\documentclass{jfm}
\usepackage{graphicx}
\usepackage{epstopdf, epsfig}
\usepackage{amsmath}
\usepackage{mathtools}
\usepackage{booktabs}
\usepackage[hidelinks]{hyperref}

\DeclareMathOperator{\erf}{erf}
\makeatletter
\ams@newcommand{\multiint}[1]{\DOTSI\protect\MultiIntegral{#1}}
\renewcommand{\MultiIntegral}[1]{%
  \edef\ints@c{\noexpand\intop
    \ifnum#1=\z@\noexpand\intdots@\else\noexpand\intkern@\fi
    \replicate{#1-2}{\noexpand\intop\noexpand\intkern@}%
    \noexpand\intop
    \noexpand\ilimits@
  }%
  \futurelet\@let@token\ints@a
}
\makeatother
\ExplSyntaxOn
\cs_new:Npn \replicate #1 #2 { \prg_replicate:nn { #1 } { #2 } }
\ExplSyntaxOff

\usepackage{color}

\shorttitle{Effect of gravity on bubble--particle collisions}
\shortauthor{T. T. K. Chan, L. Jiang and D. Krug}

\title{The effect of gravity on bubble--particle collisions in turbulence}

\author{Timothy T. K. Chan\aff{1}\corresp{\email{t.k.t.chan@utwente.nl}},
Linfeng Jiang\aff{1}
 \and Dominik Krug\aff{1,2}\corresp{\email{d.krug@aia.rwth-aachen.de}}}

\affiliation{\aff{1}Physics of Fluids Group, Max Planck Center for Complex Fluid Dynamics, and J. M. Burgers Centre for Fluids Dynamics, University of Twente, P.O. Box 217, 7500 AE Enschede, The Netherlands
\aff{2}Institute of Aerodynamics, RWTH Aachen University, Wüllnerstraße 5a, 52062 Aachen, Germany}

\begin{document}

\maketitle

\begin{abstract}
Bubble--particle collisions in turbulence are key to the froth flotation process that is widely employed industrially to separate hydrophobic from hydrophilic materials. In our previous study (Chan \textit{et al.}, \textit{J. Fluid Mech.}, vol. 959, 2023, A6), we elucidated the collision mechanisms and critically reviewed the collision models in the no-gravity limit. In reality, gravity may play a role since ultimately separation is achieved through buoyancy-induced rising of the bubbles. 
This effect has been included in several collision models, which have remained without a proper validation thus far due to a scarcity of available data. We therefore conduct direct numerical simulations of bubbles and particles in homogeneous isotropic turbulence with various Stokes, Froude, Reynolds numbers, and particle density ratios using the point-particle approximation.
Generally, turbulence enhances the collision rate compared to the pure relative settling case by increasing the collision velocity.
Surprisingly, however, for certain parameters the collision rate is lower with turbulence compared to without, independent of the history force. This is due to turbulence-induced bubble--particle spatial segregation, which is most prevalent at weak relative gravity and decreases as gravitational effects become more dominant, and reduced bubble slip velocity in turbulence. 
The existing bubble--particle collision models only qualitatively capture the trends in our numerical data. 
To improve on this, we extend the model by Dodin \& Elperin (\textit{Phys. Fluids}, vol. 14, no. 8, 2002, pp.2921-2924) to the bubble--particle case and found excellent quantitative agreement for small Stokes numbers when the history force is negligible and segregation is accounted for.
\end{abstract}

\begin{keywords}
/
\end{keywords}

\section{Introduction}\label{sec::intro_G}
Bubble--particle collisions in turbulence are central to froth flotation, which is an industrial process widely used in different contexts such as mineral extraction and wastepaper deinking. In this process, the hydrophobic target particles, which are originally suspended in a turbulent liquid mixture, are separated through colliding with bubbles, attaching to them and being floated to the surface by the bubbles' buoyancy. It is estimated that flotation has been used to treat 2 billion tons of ore and 130 million tons of paper annually in 2004 \citep{nguyen_colloidal_2004}. Given the vast scale at which flotation is employed, there is a strong incentive to understand the underlying physics, model the system, and improve efficiency.

Considering the multiscale nature of the problem, the bubble--particle interaction process is typically decomposed into the geometric collision rate, the collision efficiency, and the attachment efficiency. These refer to the bubble--particle collision rate without accounting for the local flow disturbance caused by the bubble/particle, the correction to the collision rate due to the local flow disturbance, and the probability of the particle attaching to the bubble, respectively. Since the geometric collision rate provides the baseline estimate of the bubble--particle collision rate, it will be our focus in this paper. In the literature \citep{pumir_collisional_2016}, the geometric collision rate is typically measured by the collision kernel $\Gamma$, which is related to the collision rate per unit volume by $Z_{12} = \Gamma_{12}n_1n_2$, where the subscripts 1 and 2 refer to the two colliding species and $n$ is the number density. The collision kernel can also be expressed as a normalised particle influx across a shell whose radius is
the collision distance $r_c = r_1 + r_2$ ($r_{1,2}$ are the radii of the colliding species). This shows that  $\Gamma_{12}$ is proportional to both the local particle density and the average approach velocity of the particles at collision distance. The former is characterised by the radial distribution function (RDF) at collision distance
\begin{equation}
g_{12}(r_c) = \frac{N_{pair}(r_c)/(4\pi r_c^2 \Delta r)}{N_1 N_2/V_{box}},
\end{equation}
where $N_{pair}(r_c)$ is the number of pairs within a distance of $r_c \pm \Delta r/2$, $N_{1,2}$ is the number of particles, and $V_{box}$ is the volume of the domain; while the latter is given by the effective radial approach velocity
\begin{equation}    \label{eq:Smin_G}
S^{12}_-(r_c) = -\int_{-\infty}^{0} \Delta v_r\mathrm{p.d.f.}(\Delta v_r|r_c) \mathrm{d}(\Delta v_r),
\end{equation}
where $\Delta v_r$ is the radial component of the relative velocity, which is positive when the pair separates, and $\mathrm{p.d.f.}(\Delta v_r|r_c)$ is the probability density function of $\Delta v_r$ conditioned on a pair separated by $r_c$. The collision kernel is then
\begin{equation}	\label{eq:CollisionKernelKinematic_4pi_G}
    \Gamma_{12} = 4\pi r_c^2g_{12}(r_c)S_-^{12}(r_c).
\end{equation}

{The value of the collision kernel can vary significantly within a real flotation cell since the turbulent flow field is highly inhomogeneous \citep{koh_cdf_2000}, which precludes the use of one simple expression to predict the overall collision rate. Typically, the flow is highly turbulent near the bottom of the flotation cell, where an impeller agitates the flow with the aim of promoting collisions between bubbles and particles. This is the region we focus on in the present study. Within this region, industrial-scale simulations divide the flow field into grids wherein subgrid scale models are applied to determine the local bubble--particle collision rate \citep{koh_cfd_2006}. These models are developed exclusively in the context of homogeneous isotropic turbulence (HIT), which has the advantage of having well-defined turbulence properties. We hence investigated bubble--particle collisions in HIT without gravity using the point-particle approximation in our previous study \citep{chan_bubbleparticle_2023}.} Under these conditions, the most relevant parameter is the Stokes number 
\begin{equation}    \label{eq:StokesNumber_G}
    St_i = \frac{\tau_i}{\tau_\eta} = \frac{r_i^2(2\rho_i/\rho_f + 1)}{9\nu\tau_\eta}
\end{equation}
($i = 1,2$ throughout this article and represents the colliding species). Here, $\tau_i = {r_i^2(2\rho_i/\rho_f + 1)}{/(9\nu)}$ is the particle response time {\citep{mathai_bubbly_2020}}, $\rho_i$ is the particle density, $\rho_f$ is the fluid density, $\nu$ is the kinematic viscosity, $\tau_\eta = (\nu/\varepsilon)^{1/2}$ is the Kolmogorov time scale and $\varepsilon$ is the average rate of turbulent dissipation. 
Depending on the range of $St$, \cite{chan_bubbleparticle_2023} conceptualised several collision mechanisms: at small $St$, collisions occur due to local fluid shear and the `local turnstile' effect, i.e. bubbles and particles attain opposite slip velocities in an accelerating fluid element purely due to their density differences. For collisions due to shear, \citet{saffman_collision_1956} estimated the collision kernel to be $\Gamma^{(ST)} = \sqrt{8\pi/15}r_c^3/\tau_\eta$. At intermediate $St$, non-local effects set in as the instantaneous slip velocity is increasingly determined by the path history \citep{voskuhle_prevalence_2014}. Over both small and intermediate $St$, the collision rate is reduced by the spatial segregation between bubbles and particles. Evidently, the fact that bubbles and particles have different densities causes the problem to be fundamentally different from particle--particle collisions in turbulence. Therefore, although theories of single-density particle collisions in turbulence can be a useful reference, they must be carefully considered before being extended to the bubble--particle case.

In this study, we focus on the effect of gravity, which has to be accounted for in most realistic situations and introduces an additional nondimensional parameter to the problem, namely the Froude number 
\begin{equation}
    Fr = \frac{a_\eta}{\mathfrak{g}},
\end{equation}
where $a_\eta = \eta/\tau_\eta^2$ and $\mathfrak{g}$ are the turbulence and gravitational accelerations, respectively, while $\eta$ is the Kolmogorov length scale. To investigate the effect of varying $Fr$, we conduct direct numerical simulations of bubbles and particles in HIT using the point-particle approach {\citep{maxey_equation_1983,gatignol_faxen_1983,auton_lift_1987,auton_force_1988}} and systematically vary the strength of gravity. Details on this approach will be given in \S\ref{sec::methods_G}. As the existing bubble--particle models  {for the geometric collision rate} are entirely based on concepts developed for the particle--particle case, we first review how gravity affects particle collisions in turbulence and then discuss the situation for bubble--particle collision in \S\ref{sec::theory_G}. Finally, results and conclusions are presented in \S\ref{sec::results_G} and \S\ref{sec::conclusion_G}, respectively.

\section{Theoretical Framework}    \label{sec::theory_G}
\subsection{The effect of gravity on collisions between monodispersed particles}    \label{sec::theoryMonodispersed_G}
Even for collisions between particles with the same properties (and hence the same settling velocities), gravity is known to affect the collision rate by introducing anisotropy into the system as it only acts along the vertical direction. This has implications for the preferential sampling exhibited by the particles in turbulence. Without gravity, particles with small or moderate $St$ deviate from the fluid pathlines and form clusters \citep{ireland_effect_2016,voskuhle_prevalence_2014}. This is most intuitively understood for heavy particles using the `centrifuge picture' \citep{maxey_gravitational_1987}, which suggests that the particles get expelled from regions of high rotation rate and cluster in low rotation rate and high strain rate regions. Gravity introduces an additional element to this picture by causing these particles to settle, such that they are swept by local vortical flows to preferentially sample the downward branches of vortices as they remain in the low rotation regions. As a result, their clusters are elongated in the vertical direction \citep{wang_settling_1993,bec_gravity-driven_2014}. To be precise, this picture holds only for low $St$ particles and other mechanisms that explain the elongation of particle clusters at higher $St$ have been proposed \citep{falkinhoff_preferential_2020}. Depending on the exact $St$, the RDF at collision distance can either increase or decrease \citep{ireland_effect_gavity_2016} relative to the no-gravity case.

Apart from changing the local particle concentration, gravity affects the collision velocity. With gravity, light and heavy particles rise and settle, respectively, which increases the particle slip velocity and reduces the time available for the particle to interact with the turbulent structures. For particles with $St\gtrsim 1$, this effect reduces $S_-$ as the path-history effect is weakened \citep{ireland_effect_gavity_2016}.

\subsection{The effect of gravity on collisions between particles with different response times}    \label{sec::theoryPdifftau_G}
When the colliding particles no longer have the same response time (but still have the same density), they collide even in quiescent liquid ($Fr\rightarrow 0$) by virtue of their different settling velocities $v_{T1}$ and $v_{T2}$, which are obtained by balancing drag with buoyancy in the vertical direction. In this `relative settling' limit, the collision kernel is given by \citep{pumir_collisional_2016}
\begin{equation}    \label{eq:relativeSettlingCollK}
    \Gamma^{(G)}_{12} = 4\pi r_c^2 S_-^{(G)} = \pi r_c^2 |v_{T1} - v_{T2}|,
\end{equation}
where the prefactor in the last equality results from the facts that collisions only occur on a hemisphere (contributing a 1/2 prefactor to $S_-^{(G)}$) and that only the radial component of the relative velocity is considered (contributing the other 1/2 prefactor to $S_-^{(G)}$).

In the intermediate $Fr$ regime, the relative settling mechanism and turbulence are active at the same time. 
Several models which originally consider the no-gravity case at the very small or very large $St$ limits (refer to \citet{chan_bubbleparticle_2023} for details) have been developed to incorporate relative settling. At the small $St$ limit, \citet{saffman_collision_1956} approximated the p.d.f. of the relative velocity by a Gaussian distribution and assumed that relative settling changed the variance of the p.d.f.. However, the resulting expression fails to converge to their no-gravity limit as acknowledged by the original authors, and gravity is included in all directions since the variance was incorrectly assumed to be isotropic. \citet{dodin_collision_2002} resolved these issues by rigorously deriving the collision kernel in the small $St$ limit to give
\begin{equation}
    \Gamma_{12}^{(DE)} = \sqrt{8\pi}r_c^2\sigma \bigg[\frac{\sqrt{\pi}}{2}\bigg(c + \frac{1}{2c}\bigg)\erf c + \frac{\exp(-c^2)}{2}\bigg],
\end{equation}
where
$c = |\tau_2 - \tau_1|\mathfrak{g}/(\sqrt{2}\sigma)$ and $\sigma^2 = r_c^2\varepsilon/(15\nu) + 1.3(\tau_2 - \tau_1)^2\varepsilon^{3/2}/\nu^{1/2}$.
Crucially, they assumed that relative settling is captured by a shift in the mean vertical velocity while the variance remains unaffected. Consequently, their expression is consistent with \citet{saffman_collision_1956} in the no-gravity limit. At the very large $St$ limit, where particles can be modelled as kinetic gases with normally distributed velocity components, \citet{abrahamson_collision_1975} similarly accounted for gravity by shifting the vertical velocity p.d.f. of each species by the mean settling velocity so that
\begin{eqnarray}    \label{eq:AbrahamsonGravityIntegral}
    \Gamma_{12} &=& \pi r_c^2 v_c\nonumber\\ &=& \frac{\pi r_c^2}{(2\pi v_1'^2)^{3/2}(2\pi v_2'^2)^{3/2}}\multiint{6}\limits_{\mathrm{all\,space}} \sqrt{(v_{1x}-v_{2x})^2 + (v_{1y}-v_{2y})^2 + (v_{1z}-v_{2z})^2}\nonumber\\
    && \times\exp{\bigg[\sum_{i = 1}^{2}-\frac{v_{ix}^2 + v_{iy}^2 + (v_{iz} - v_{Ti})^2}{2v_i'^2} \bigg]} \mathrm{d}v_{1x}\mathrm{d}v_{1y}\mathrm{d}v_{1z}\mathrm{d}v_{2x}\mathrm{d}v_{2y}\mathrm{d}v_{2z},
\end{eqnarray}
where $v_{x,y}$ and $v_{z}$ are the horizontal and the vertical particle velocity components, respectively. \citet{abrahamson_collision_1975} evaluated the integral in (\ref{eq:AbrahamsonGravityIntegral}) analytically and reported that
\begin{equation}    \label{eq:AbrahamsonGravity_original}
    \frac{v_c}{\chi} = \frac{2^{3/2}}{\sqrt{\pi}}\exp\bigg(-\frac{\alpha^2}{2}\bigg) + \bigg(\alpha + \frac{1}{\alpha}\bigg)\erf{\bigg(\frac{\alpha}{\sqrt{2}}\bigg)},
\end{equation}
where
\begin{equation}    \label{eq:alphaChi_def}
    \alpha = \frac{v_{T2}-v_{T1}}{\chi}, \quad \chi = \sqrt{v'^{2}_1 + v'^{2}_2},
\end{equation}
and $v'^2_i$ is the mean-square single-component particle velocity. Unfortunately, the integration was not performed correctly so (\ref{eq:AbrahamsonGravity_original}) does not reduce to the no-gravity case ($\Gamma_{12} = \sqrt{8\pi}r_c^2\sqrt{v'^{2}_1 + v'^{2}_2}$) when $\alpha \to 0$ because $\lim_{\alpha\to0} \erf(\alpha)/\alpha$ does not converge to 0, as has already been pointed out by \citet{kostoglou_generalized_2020,kostoglou_critical_2020}. Instead, we integrate (\ref{eq:AbrahamsonGravityIntegral}) numerically and obtain the best-fit result
\begin{equation}    \label{eq:AbrahamsonCollKBasic}
    \Gamma_{12} = \pi r_c^2 v_c
\end{equation}
where
\begin{equation}    \label{eq:AbrahamsonGravIntegrationFit}
\frac{v_c}{\chi} = 
\left\{
\begin{aligned}
1.6,     \quad&\text{for $\alpha < 0.1$}\\
-0.0188\alpha^3+0.2174\alpha^2+0.1073\alpha+1.5552,\quad&\text{for $0.1\leq \alpha\leq 5$}\\
\alpha + (1/\alpha),\quad&\text{for $\alpha>5$}
\end{aligned}
\right.
\end{equation}
and $\alpha$ and $\chi$ are given by (\ref{eq:alphaChi_def}). Note that the coefficients for the $\alpha \in [0.1,5]$ case of (\ref{eq:AbrahamsonGravIntegrationFit}) are slightly different from the ones reported in \citet{kostoglou_generalized_2020}, presumably due to a typographical error (see Appendix \ref{sec::AbrahamsonincludeGravity}).

{None of the above models account for the intricate interactions between the effects of turbulence and gravity, as pointed out by \citet{woittiez_combined_2009}. This is particularly important for bidispersed particles whose motions are correlated. Essentially, the different settling speeds means that the two species do not have the same amount of time to interact with the fluid locally. This leads to a further decorrelation of their concentration fields \citep{woittiez_combined_2009,dhariwal_small-scale_2018}, as well as increased collision velocity as particles attain high values of acceleration in the horizontal direction \citep{parishani_effects_2015,dhariwal_small-scale_2018}, both of which affects the collision rate. Despite the limitations of the above models, we introduce them here as they form the basis for the existing bubble--particle collision models.}

\subsection{The effect of gravity on bubble--particle collisions}   \label{sec::theory_bpG}
With gravity, bubbles rise and particles settle due to their relative density, meaning $v_{T1}$ and $v_{T2}$ have opposite signs in (\ref{eq:relativeSettlingCollK}) so relative settling is enhanced. 
Existing models for bubble--particle collisions are almost all direct extensions of those discussed in \S\ref{sec::theoryPdifftau_G}. At the high $St$, `kinetic gas-like' limit, \citet{bloom_structure_2002} adapted ({\ref{eq:AbrahamsonGravityIntegral}} -- \ref{eq:alphaChi_def}) by simply replacing the settling velocities with the bubble (particle) rising (settling) velocity, meaning
\begin{equation}    \label{eq:alphaChi_defBH}
    \alpha^{(BH)} = \frac{|v_{Tb}|+|v_{Tp}|}{\chi^{(BH)}} \quad \mathrm{and} \quad \chi^{(BH)} = \sqrt{v'^{2}_b + v'^{2}_p},
\end{equation}
where the subscripts $b$ and $p$ denote `bubble' and `particle', respectively. {Since \citet{bloom_structure_2002} incorrectly used expressions for slip velocities in lieu of $v'_b$ and $v'_p$ as pointed out in \citet{chan_bubbleparticle_2023}, in this work we employ the expression given by \citet{abrahamson_collision_1975} in (\ref{eq:alphaChi_defBH}) for both $v'_b$ and $v'_p$.} For intermediate $St$, \citet{ngo-cong_isotropic_2018} first modelled the zero-gravity bubble--particle velocity correlation following the approach of \citet{yuu_collision_1984}, such that the root mean square (r.m.s.) of the single-component bubble--particle relative velocity without gravity is $\sigma_{\Delta v}^{(NC)} = [(A_b + A_p - 2B)u'^2 + (A_br_b^2 + A_pr_p^2 + 2Br_pr_b)\varepsilon/(9\nu)]^{1/2}$, where $u'$ is the single-component r.m.s. fluid velocity fluctuation, and $A_i$ and $B$ are coefficients that depend on particle properties and the fluid Lagrangian integral time scale. They then included gravity using (\ref{eq:AbrahamsonGravity_original}) with
\begin{equation}    \label{eq:alphaChi_defNC}
    \alpha^{(NC)} = \frac{|v_{Tb}|+|v_{Tp}|}{\chi^{(NC)}} \quad \mathrm{and} \quad \chi^{(NC)} = \sigma_{\Delta v}^{(NC)}.
\end{equation}
We point out that (\ref{eq:AbrahamsonGravity_original}) is derived from (\ref{eq:AbrahamsonGravityIntegral}) which assumes uncorrelated particle velocities. The use of (\ref{eq:AbrahamsonGravity_original}) at intermediate $St$, where the correlation between particle velocities remains finite, is therefore inconsistent. 
In addition, both \citet{bloom_structure_2002} and \citet{ngo-cong_isotropic_2018} use the wrong integration result (\ref{eq:AbrahamsonGravity_original}), such that these models also do not converge to the correct no-gravity limit. Expressions that have been corrected for this error are given by (\ref{eq:AbrahamsonCollKBasic}) and (\ref{eq:AbrahamsonGravIntegrationFit}), along with (\ref{eq:alphaChi_defBH}) for \citet{bloom_structure_2002} and (\ref{eq:alphaChi_defNC}) 
for \citet{ngo-cong_isotropic_2018}. For small $St$, a consistent model for bubble--particle collisions in turbulence with gravity is missing to date. We extend the theory by \citet{dodin_collision_2002} to the bubble--particle case (see Appendix \ref{sec::extDodinElperin_bpG} for details) and include a drag correction $f_i$ ($f_i = 1$ for Stokes drag) to obtain
\begin{equation}
    \Gamma_{bp}^{(DEX)} = \sqrt{8\pi}r_c^2\sigma_{\Delta vr}^{(DEX)} \bigg[\frac{\sqrt{\pi}}{2}\bigg(c + \frac{1}{2c}\bigg)\erf c + \frac{\exp(-c^2)}{2}\bigg],
\end{equation}
where
\begin{equation}
    \sigma_{\Delta vr}^{(DEX)} = \sqrt{\frac{r_c^2\varepsilon}{15\nu} + 1.3\bigg(\frac{\beta_b\tau_b}{\langle f_b\rangle} - \frac{\beta_p\tau_p}{\langle f_p\rangle}\bigg)^2\frac{\varepsilon^{3/2}}{\nu^{1/2}}}
\end{equation}
is the r.m.s. of $\Delta v_r$ in zero gravity,
\begin{equation}
    c = \frac{|\beta_b\tau_b/\langle f_b\rangle - \beta_p\tau_p/\langle f_p\rangle|\mathfrak{g}}{\sqrt{2}\sigma_{\Delta vr}^{(DEX)}}
\end{equation}
is the ratio of the relative settling velocity to the turbulence-induced radial collision velocity (up to a constant), and $\langle\cdot\rangle$ denotes averaging. This expression incorporates three collision mechanisms at small $St$: collision due to local fluid shear, local turnstile mechanism and relative settling. We note that \citet{kostoglou_generalized_2020} proposed a model which considers a bubble in a swarm of tracers. However, since it does not decompose the collision process into the geometric collision rate and the collision efficiency, we do not further consider it in this study.

While several models for bubble--particle collisions in turbulence that account for gravity are available, none of them have been directly tested. Even in zero gravity, the model predictions are hugely different from simulation results which underscores the lack of understanding and the richness of the physics of bubble--particle collisions \citep{chan_bubbleparticle_2023}. 
The only direct numerical study on bubble--particle geometric collisions with gravity in HIT that we are aware of does not vary $St$, $Fr$ and $\Rey_\lambda$ independently \citep{wan_study_2020}. The goal this work is therefore to systematically investigate the effect of $St$, $Fr$ and $\Rey_\lambda$ for bubble--particle collisions through simulations in order to reveal the underlying physical picture and assess the available models.

\section{Methods}   \label{sec::methods_G}
\begin{figure}
    \centering
    \includegraphics{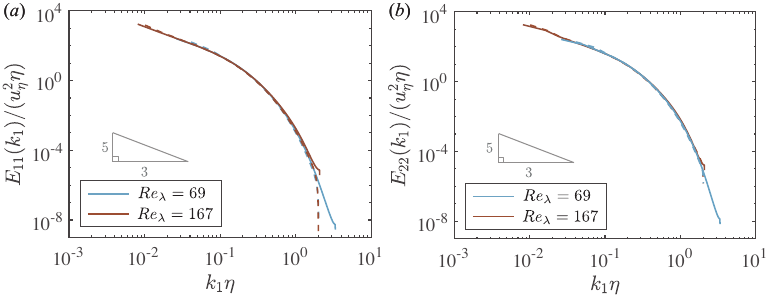}
    \caption{The (a) longitudinal and (b) transverse energy spectra. The dashed lines show data from \citet{jimenez_structure_1993} for $\Rey_\lambda = 62.7$ (in blue) and $170.2$ (in brown).}
    \label{fig:pspec_G}
\end{figure}

The simulations are performed using the same fluid and particle solvers as \citet{chan_bubbleparticle_2023}. Hence we only provide a brief summary here for completeness. For the background turbulence, we run direct numerical simulations of HIT using a finite-difference solver (second order in space, third order in time) to solve the incompressible Navier--Stokes and continuity equations
\begin{subeqnarray}
	\frac{\mathrm{D}\boldsymbol{u}}{\mathrm{D}t} &=& -\frac{1}{\rho_f}\nabla P + \nu \nabla^2\boldsymbol{u} + \boldsymbol{f_\Psi},\\
	\nabla\boldsymbol{\cdot}\boldsymbol{u} &=& 0,
\end{subeqnarray}
where $\mathrm{D}/\mathrm{D}t$ is the material derivative following a fluid element, $\boldsymbol{u}$ is the fluid velocity, $t$ is the time, and $P$ is the pressure. The fluid is forced randomly at the largest scales $\boldsymbol{f_\Psi}$ using the scheme by \citet{eswaran_examination_1988} to generate turbulence with $\Rey_\lambda = 69$ and $167$. Other turbulence statistics are displayed in table \ref{tab:turbStat_G}. We also show in figure \ref{fig:pspec_G} that the power spectrum agrees excellently with that of \citet{jimenez_structure_1993}.

For the bubbles and particles, we approximate them using the point-particle approach {\citep[for a historical overview of this approach, see][]{michaelides_hydrodynamic_2003} which implies that both species are modelled as rigid spheres. This is reasonable even for the bubbles since the Weber number based on their rise velocity at $(St_b, 1/Fr, Re_\lambda) = (11, 4.4, 64)$ is O(0.1) \citep{jiang_how_2024}. As described later in this section, this work focuses on bubbles with lower values of $St_b$, such that the bubble radii and rise velocities are smaller at comparable values of $1/Fr$, thus bubble deformation is even less significant. Under the point-particle approach \citep{maxey_equation_1983,gatignol_faxen_1983,auton_force_1988}, }the net force on the particle is the sum of the drag force, the pressure gradient force, the added mass force, {buoyancy, lift \citep{auton_lift_1987}, the history force, and Faxen terms. We initially neglect the history force and Faxen terms to allow fair comparison with the models introduced in \S\ref{sec::theory_bpG}.} A finite-difference solver is used to solve
\begin{eqnarray}\label{eq:PointParticleEoM_G}
\frac{4}{3}\pi r_i^3\rho_i\frac{\mathrm{d}\boldsymbol{v_i}}{\mathrm{d}t} & = & 6\pi\mu r_if_i(\boldsymbol{u}-\boldsymbol{v_i}) + \frac{4}{3}\pi r_i^3\rho_f\frac{\mathrm{D}\boldsymbol{u}}{\mathrm{D}t} + \frac{2}{3}\pi r_i^3\rho_f\bigg(\frac{\mathrm{D}\boldsymbol{u}}{\mathrm{D}t} - \frac{\mathrm{d}\boldsymbol{v_i}}{\mathrm{d}t}\bigg)\nonumber\\ &&
+ \frac{4}{3}\pi r_i^3(\rho_f - \rho_i)\mathfrak{g}\boldsymbol{e_z} - \frac{2}{3}\rho_f\pi r_i^3(\boldsymbol{v_i} - \boldsymbol{u})\times\boldsymbol{\omega},
\end{eqnarray}
where $\boldsymbol{v_i}$ is the particle velocity, $\mu=\nu\rho_f$ is the absolute viscosity, $\boldsymbol{e_z}$ is the unit vector pointing vertically upwards, $\boldsymbol{\omega}$ is the vorticity vector, and $f_i = 1 + 0.169Re_i^{2/3}$ is the correction factor that accounts for nonlinear drag due to finite bubble or particle Reynolds number $Re_i = 2r_i|\boldsymbol{u - v_i}|/\nu$ and implies a no-slip boundary condition \citep{nguyen_colloidal_2004}. $r_i$ is determined from (\ref{eq:StokesNumber_G}), while $\boldsymbol{u}$, $\mathrm{D}\boldsymbol{u}/\mathrm{D}t$ and $\boldsymbol{\omega}$ at the particle positions are evaluated using tri-cubic Hermite spline interpolation with a stencil of four points per direction \citep{van_hinsberg_optimal_2013}. 
We employ a lift coefficient $C_L$ of 1/2, which strictly speaking is only valid for clean bubbles with very large $Re_b$ in simple shear flows \citep{auton_lift_1987,legendre_lift_1998}, and acknowledge that the actual value depends on the type of flow and $Re_b$ \citep{legendre_lift_1998,magnaudet_aspects_1998}. Nonetheless, we use $C_L = 1/2$ following \citet{mazzitelli_lagrangian_2004} who simulated similar bubbles in HIT. {Although we initially neglect the history force, we recognise that it may play a noticeable role in reality for our tested bubble and particle density ratios \citep{daitche_role_2015,olivieri_effect_2014}. We therefore run additional cases in \S \ref{sec::historyForce_G} with the history force
\begin{equation}    \label{eq:bassetHistoryForce}
    \boldsymbol{F_{hist}} = 6r_i^2\rho_f\sqrt{\pi\nu}\int_{0}^{t} \frac{\frac{\mathrm{d} (\boldsymbol{u}(\tau) - \boldsymbol{v_i}(\tau))}{\mathrm{d}\tau}}{\sqrt{t-\tau}}\mathrm{d}\tau
\end{equation}
added to the right hand side of (\ref{eq:PointParticleEoM_G}). $\boldsymbol{F_{hist}}$ is computed using the semi-implicit scheme by \citet{van_hinsberg_efficient_2011}, where a window length of 5 time steps is chosen and the tail of the integrand is modelled with 10 exponential functions. Hence, the results are comparable to a simple window method \citep{dorgan_efficient_2007} where the integration is truncated to the last 500 time steps \citep{van_hinsberg_efficient_2011}. This corresponds to approximately $9\tau_\eta$, which \citet{calzavarini_impact_2012} found is long enough for the integrand to decay to negligible values in HIT for their parameters.}

{In all our simulations, the} collisions are treated using the `ghost collision' scheme as it is consistent with the models described in \S\ref{sec::theory_G} \citep{wang_collision_1998}. This means particles can overlap and a collision is registered every time an approaching pair of particles reaches the collision distance $r_c$, which is set to $r_b + r_p$ for all species. We eliminate wall effects by implementing periodic boundary conditions in all directions and verified that the domain size $L_{box} = 1$ is sufficiently large to minimise periodicity effects (see Appendix \ref{sec::domainSize_G}). {The time step $\Delta t \leq \tau_\eta/55$ is sufficiently small such that the CFL number $  \leq 1$ (apart from a few time steps for the cases in \S\ref{sec::rhopInf_G} and \S\ref{sec::historyForce_G}) and the point-particle statistics are no longer sensitive to $\Delta t$ \citep{ruth_effect_2021}.} We refer the reader to \citet{chan_bubbleparticle_2023} for details of the solvers and the collision detection algorithm.

\begin{table}
  \begin{center}
  \begin{tabular}{lccccccccc}
      $\Rey_\lambda$  & $\mathcal{N}$   &   $\overline{\varepsilon}$ & $\eta$ & $k_{max}\eta$ & $u_\eta$ & $u'/u_\eta$ & $u_x'/u_y'$ & $T_L/\tau_\eta$ & $N_{b,p}$\\[3pt]
       69 & $256^3$ & 51.9 & 0.0042 & 3.4 & 0.60 & 4.2 & 0.99 & 18 & 10000\\
       167 & $512^3$ & 356 & 0.0013 & 2.1 & 0.77 & 6.6 & 0.95 & 43 & $77700-140000$\\
  \end{tabular}
  \caption{Statistics of the homogeneous isotropic turbulence: the grid size ($\mathcal{N}$), pseudo-dissipation ($\overline{\varepsilon}$), Kolmogorov length scale ($\eta$), maximum wavenumber ($k_{max}$), Kolmogorov velocity ($u_\eta$) scale, root-mean-square velocity fluctuations ($u'$), large-scale isotropy ($u_x'/u_y'$), and the large eddy turnover time ($T_L$) relative to the Kolmogorov time scale ($\tau_\eta$). $N_{b,p}$ are the numbers of bubbles and particles respectively. {The simulations including the history force have the same parameters as the $Re_\lambda = 167$ case except for a lower number of bubbles and particles $N_{b,p} = 50000$.}}
  \label{tab:turbStat_G}
  \end{center}
\end{table}

We investigate bubbles ($\rho_b/\rho_f = 1/1000$) and particles ($\rho_p/\rho_f = 5$) with {$St$ of 0.1, 0.5, 1, 2 and 3, and $1/Fr$ ranging from 0.01 to 10. For the bubbles, these values of $St$ correspond to $r_b/\eta = 0.9$, 2.2, 3.0, 4.2 and 5.2, respectively; and it should be noted that finite-size effects, that are not represented in our computations, may become relevant in particular at the larger values of $St$. Nonetheless, we believe that our approach still allows us to examine the general trend with increasing St and to assess the collision models introduced in \S\ref{sec::theory_bpG}, which also do not include these effects. 
We consider only $St_b = St_p$ to limit the parameter space. While this is not the most general case, it is physically relevant especially in the context of flotation of fine particles, where small bubbles are preferred \citep{ahmed_effect_1985,miettinen_limits_2010,farrokhpay_flotation_2021}. The simulation is conducted as follows: bubbles and particles, $10000-140000$ per species,} are injected in the simulation domain at random positions after the flow has reached statistical stationarity. We then monitor the particle position p.d.f., particle velocity p.d.f. and the ensemble-average rise/settling velocities to identify transients. Once the transient states have passed, we collect statistics over at least $12.4$ large eddy turnover times ($T_L$) at a minimum sampling frequency of once per $0.06 T_L$.

\section{Results}   \label{sec::results_G}
\subsection{Collision kernel}
\begin{figure}
    \centering
    \includegraphics{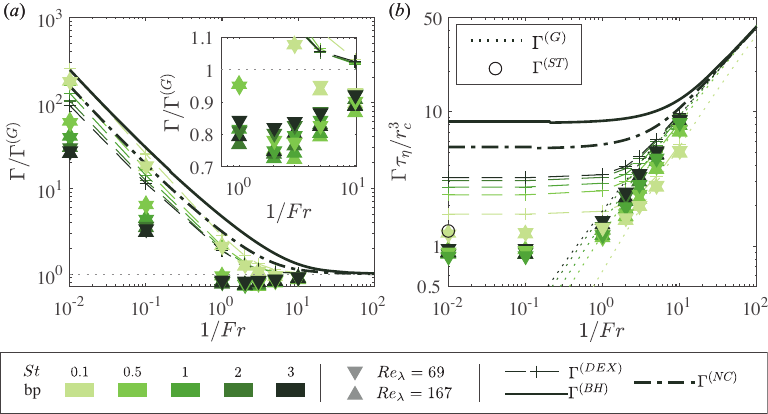}
    \caption{(\textit{a}) The bubble--particle collision kernel normalised by the result for the relative settling case in still fluid. The inset shows a close-up view of the region where $1/Fr \geq 1$. (\textit{b}) The bubble--particle collision kernel normalised by $r_c^3/\tau_\eta$. $\Gamma^{(BH)}$ and $\Gamma^{(NC)}$ (only shown for $St = 3$) are computed using the corrected best-fit (\ref{eq:AbrahamsonGravIntegrationFit}). The model predictions are for $\Rey_\lambda = 167$.}
    \label{fig:collK_G}
\end{figure}

Figure \ref{fig:collK_G}(\textit{a}) presents results for the bubble--particle collision kernel normalised by the relative settling case in still fluid $\Gamma^{(G)}$, wherein the terminal velocities $v_{Ti}$ are obtained by balancing buoyancy with the nonlinear drag term as shown by (\ref{eq:PointParticleEoM_G}). At small $1/Fr$ (weak gravity), $\Gamma/\Gamma^{(G)}$ is large since bubbles and particles rise and settle very slowly such that $\Gamma^{(G)}$ is small and the collision rate is dominated by turbulence mechanisms. In this regime, $\Gamma/\Gamma^{(G)}$ reduces with increasing $St$ primarily because the rise and settling speeds $|v_{Ti}|$ are proportional to $St$. As $1/Fr$ increases, bubbles and particles rise and settle faster, enhancing the role of relative settling in the overall collision rate as reflected by a decreasing $\Gamma/\Gamma^{(G)}$ such that at large $1/Fr$, $\Gamma \approx \Gamma^{(G)}$. At close inspection though, $\Gamma < \Gamma^{(G)}$ when $1/Fr \gtrsim 5$ for $St = 0.1$ and $1/Fr \gtrsim 1$ for $St \geq 0.5$ as shown in the inset. This implies that, counter-intuitively, the presence of turbulence decreases the collision rate below the relative settling case in this regime. This is attributed to the nonuniform bubble--particle spatial distribution and their resulting segregation, and to the reduced bubble {slip} velocity due to nonlinear drag, as will be discussed further in \S\ref{sec::bpDistribution_G} and \S\ref{sec::collisionVelocity_G}. As the segregation does not monotonically depend on $St$, there is no simple dependence between $\Gamma/\Gamma^{(G)}$ and $St$ in this regime. Throughout the tested range of $1/Fr$ {and $\Rey_\lambda$}, $\Gamma/\Gamma^{(G)}$ is not sensitive to $\Rey_\lambda$.

All the models considered successfully capture the general decreasing trend of $\Gamma/\Gamma^{(G)}$ and the relative settling limit $\Gamma \to \Gamma^{(G)}$ at $1/Fr \to \infty$. However, the model predictions remain above $\Gamma^{(G)}$ over the entire range of $1/Fr$ even though the nonlinear drag expression has already been included. This is because they assume a uniform bubble--particle distribution and employ the bubble rise velocity in still fluid. As mentioned at the end of the last paragraph, this is not true from the simulations and will be elaborated on in \S\ref{sec::bpDistribution_G} and \S\ref{sec::collisionVelocity_G}.

We now normalise the collision kernel by $\tau_\eta/r_c^3$, i.e. the scaling of $\Gamma^{(ST)}$, in figure \ref{fig:collK_G}(\textit{b}) to better examine the turbulence-to-relative settling transition. When $1/Fr \lesssim 0.1$, bubble and particle dynamics are turbulence-dominated and the results are consistent with \citet{chan_bubbleparticle_2023}: the collision kernel coincides with $\Gamma^{(ST)}$ at small $St$ and $\Gamma^{(NC)}$ and $\Gamma^{(BH)}$ overpredict the collision kernel. We stress, however, that the agreement with $\Gamma^{(ST)}$ is merely a coincidence since the model does not account for bubble--particle segregation as discussed in \citet{chan_bubbleparticle_2023}. When $1/Fr$ increases, the effect of gravity on the collision kernel becomes noticeable and causes it to increase towards $\Gamma^{(G)}$. In the tested parameter range, $\Gamma$ is only weakly sensitive to $\Rey_\lambda$. 
Among the models, $\Gamma^{(DEX)}$ best matches the data for the parameters investigated. This is expected since it has been developed for the small $St$ limit and has been extended to the bubble--particle case to also incorporate the local turnstile mechanism. On the other hand, \citet{bloom_structure_2002} is applicable for the large $St$ kinetic gas limit which is inaccessible with the point-particle approach. Nonetheless, even for the $St = 0.1$ case $\Gamma^{(DEX)}$ does not quantitatively agree with the data, presumably due to preferential sampling as will be examined in the following section.

\subsection{Bubble--particle distribution}  \label{sec::bpDistribution_G}
\begin{figure}
    \centering
    \includegraphics{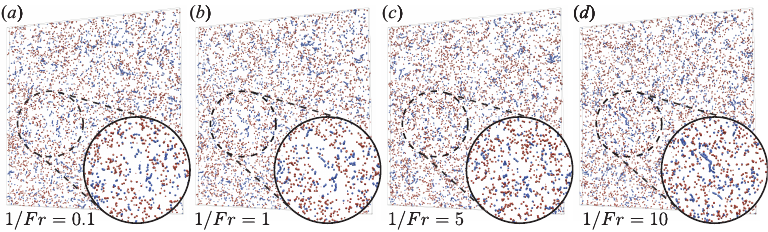}
    \caption{Instantaneous snapshots of bubbles (blue) and particles (red) in a slice with width $\times$ height $\times$ depth = $L_{box} \times L_{box} \times 20\eta$ with $\Rey_\lambda = 167$, $St = 1$, and (\textit{a}) $1/Fr = 0.1$, (\textit{b}) $1/Fr = 1$, (\textit{c}) $1/Fr = 5$ and (\textit{d}) $1/Fr = 10$. Gravity is directed vertically downwards. The sizes of the bubbles and particles are not to scale.}
    \label{fig:segregation_G}
\end{figure}

\begin{figure}
    \centering
    \includegraphics{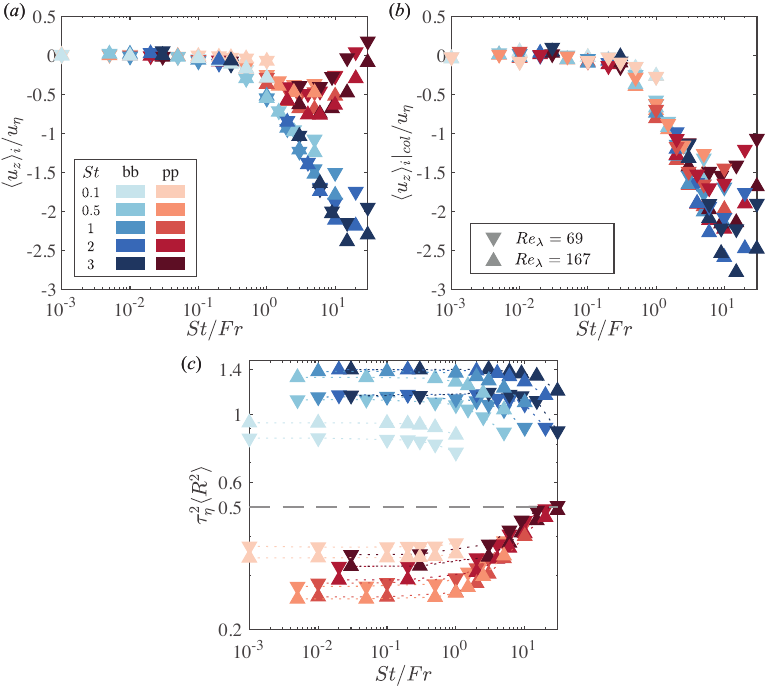}
    \caption{(\textit{a}) Mean vertical fluid velocity at bubble and particle positions plotted against $St/Fr$. (\textit{b}) Same as (\textit{a}) but conditioned on pairs with $r\in[r_c-0.1\eta,r_c + 0.1\eta]$. (\textit{c}) The norm of the rotation rate of the flow at bubble and particle positions. The lines are guides for the eye.}
    \label{fig:fluidVelPpos}
\end{figure}

In zero gravity, bubbles and particles with small or moderate $St$ preferentially sample different flow regions in HIT due to different relative densities and form respective clusters at regions of high and low rotation rates, meaning they segregate \citep{calzavarini_quantifying_2008-1}. This effect has been shown to be most pronounced at $St = 1$ \citep{chan_bubbleparticle_2023}. To examine the effect of gravity on the spatial distribution of bubbles and particles, figure \ref{fig:segregation_G} displays the instantaneous snapshots of the simulation with $St = 1$ over a range of $1/Fr$. As $1/Fr$ becomes larger, the overlap between bubble and particle positions increases. 
However, we do not observe a clear elongation of the bubble and particle clusters in the vertical direction in contrast to other studies of heavy particles in turbulence \citep{ireland_effect_gavity_2016,falkinhoff_preferential_2020,bec_gravity-driven_2014}. This is because these studies employed a higher $\rho_p$ which enhances the overall effect of gravity. To more quantitatively investigate the locations sampled by bubbles and particles, we consider the theory by \citet{bec_gravity-driven_2014}, which predicts that a negative horizontal divergence of $\boldsymbol{v_p}$ occurs on average in downward flows when particles are settling in turbulence, i.e. `preferentially sweeping' \citep{wang_settling_1993}. Extending this reasoning to buoyant bubbles would mean that bubbles also cluster in downflow regions. Furthermore, the lift force pulls rising bubbles toward the downward branches of vortices \citep{mazzitelli_effect_2003}. Collectively, this implies bubbles and particles sample downward flows in turbulence due to gravity. We test this hypothesis by plotting the mean vertical fluid velocity at bubble and particle positions in figure \ref{fig:fluidVelPpos}(\textit{a}) and show that this is indeed the case for $St/Fr \gtrsim 10^{-1}$, with bubbles sampling regions with stronger downflow than particles. Additionally, the figure shows that $\langle u_z \rangle_i$ collapses especially for $St \geq 0.5$ when plotted against $St/Fr$, which echos the finding by \citet{good_settling_2014} that $St/Fr$ is a key indicator of different particle settling regimes. To single out the colliding bubbles and particles, we condition the fluid velocity on pairs that are close to the collision distance in figure \ref{fig:fluidVelPpos}(\textit{b}). We observe that the colliding pairs also preferentially sample downflow regions, and $\langle u_z \rangle_i|_{col}$ essentially takes the unconditioned bubble values $\langle u_z \rangle_b$. This suggests that the bubble--particle collisions occur mainly at the unconditioned bubble positions such that the collision rate depends on the location of the particle clusters relative to the bubbles. To further examine this, we plot the norm of the rotation rate $\langle R^2 \rangle$ at bubble and particle positions in figure \ref{fig:fluidVelPpos}(\textit{c}). When gravity is weak, bubbles and particles preferentially sample regions with high and low $\langle R^2 \rangle$ {values}, respectively. As gravity becomes stronger, {the values of} both particle and (to a lesser extent) bubble $\langle R^2 \rangle$ begin to return to the tracer limit of 0.5, which indicates a change in the clustering location and/or bubbles and particles cluster less. 

\begin{figure}
    \centering
    \includegraphics{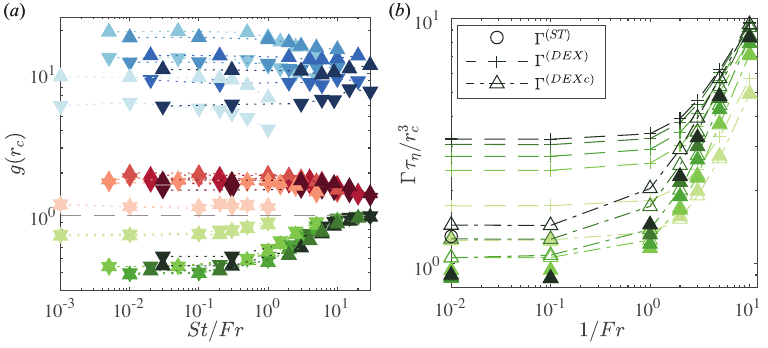}
    \caption{(\textit{a}) The radial distribution function at collision distance $g(r_c)$. Only 25\% of all the bubbles were considered when computing $g_{bb}(r_c)$. (\textit{b}) The collision kernel and the prediction of the extended \citet{dodin_collision_2002} model after compensating for $g(r_c)$ plotted against $1/Fr$ for $\Rey_\lambda = 167$. The lines are guides for the eye. The symbols and colour scheme follows figures \ref{fig:collK_G} and \ref{fig:fluidVelPpos}.}
    \label{fig:RDF_G}
\end{figure}

To directly measure the overall effect of preferential sampling on the collision kernel, we compute the radial distribution function at collision distance $g(r_c)$, which reflects the probability of finding a particle at a distance $r_c$ from another particle relative to a uniform distribution. In short, $g(r_c) < 1$ and $>1$ indicate segregation and clustering, which respectively either decreases or increases the collision rate based on (\ref{eq:CollisionKernelKinematic_4pi_G}). Figure \ref{fig:RDF_G}(\textit{a}) shows $g(r_c)$ as a function of $St/Fr$. Focusing on the lowest $1/Fr$ case when gravity is still weak, bubbles and particles form their own clusters so $g_{bb}(r_c)$ and $g_{pp}(r_c)>1$. However, since bubbles and particles individually cluster at different regions of the flow owing to their different relative densities, they segregate and $g_{bp}(r_c)<1$. The segregation is strongest at $St = 1$ when $g_{bp}(r_c)$ reaches its lowest value, which is consistent with the no gravity case \citep{chan_bubbleparticle_2023}. As $1/Fr$ increases, {the extent of segregation reduces and eventually reaches $g_{bp}(r_c) = 1$.} We note that $g_{bp}(r_c)$ for $St \geq 0.5$ collapses when plotted against $St/Fr$ as shown in figure \ref{fig:RDF_G}(\textit{a}). This may be because particle clusters are affected by preferential sampling \citep{wang_settling_1993}, which scales as $St/Fr$ as shown in figure \ref{fig:fluidVelPpos}(\textit{a}). {Curiously, neither $g_{bb}(r_c)$ nor $g_{pp}(r_c)$ reaches 1 even at the largest tested value of $St/Fr$, which indicates that reduced bubble or particle clustering is not the primary cause of the reduced segregation. On closer inspection, }$g_{bp}(r_c)$ for $St \geq 0.5$ already deviates noticeably from the $1/Fr \rightarrow 0$ limit between $10^{-1} \lesssim St/Fr \lesssim 10^{0}$ despite $g_{pp}(r_c)$ and $g_{bb}(r_c)$ remaining mostly constant, i.e. bubbles and particles still cluster as strongly as before. From figure \ref{fig:fluidVelPpos}(\textit{c}), in this range of $St/Fr$, the {value of the} bubble $\langle R^2 \rangle$ already decreases for $St = 0.5$ and the {value of the} particle $\langle R^2 \rangle$ already increases for $St \geq 1$. This suggests the bubbles and particle clusters migrate closer together which would increase $g_{bp}(r_c)$. For the tested parameters, $g_{bp}(r_c)$ is not sensitive to $\Rey_\lambda$. 
To account for segregation in the model by \citet{dodin_collision_2002}, we multiply $g_{bp}(r_c)$ to $\Gamma^{(DEX)}$ to obtain $\Gamma^{(DEXc)}$, which is plotted in figure \ref{fig:RDF_G}(\textit{b}). As anticipated, $\Gamma^{(DEXc)}$ agrees excellently with the simulation data at the lowest $St$, indicating that the model correctly represents the relevant physics when corrected for the effect of segregation.

\begin{figure}
    \centering
    \includegraphics{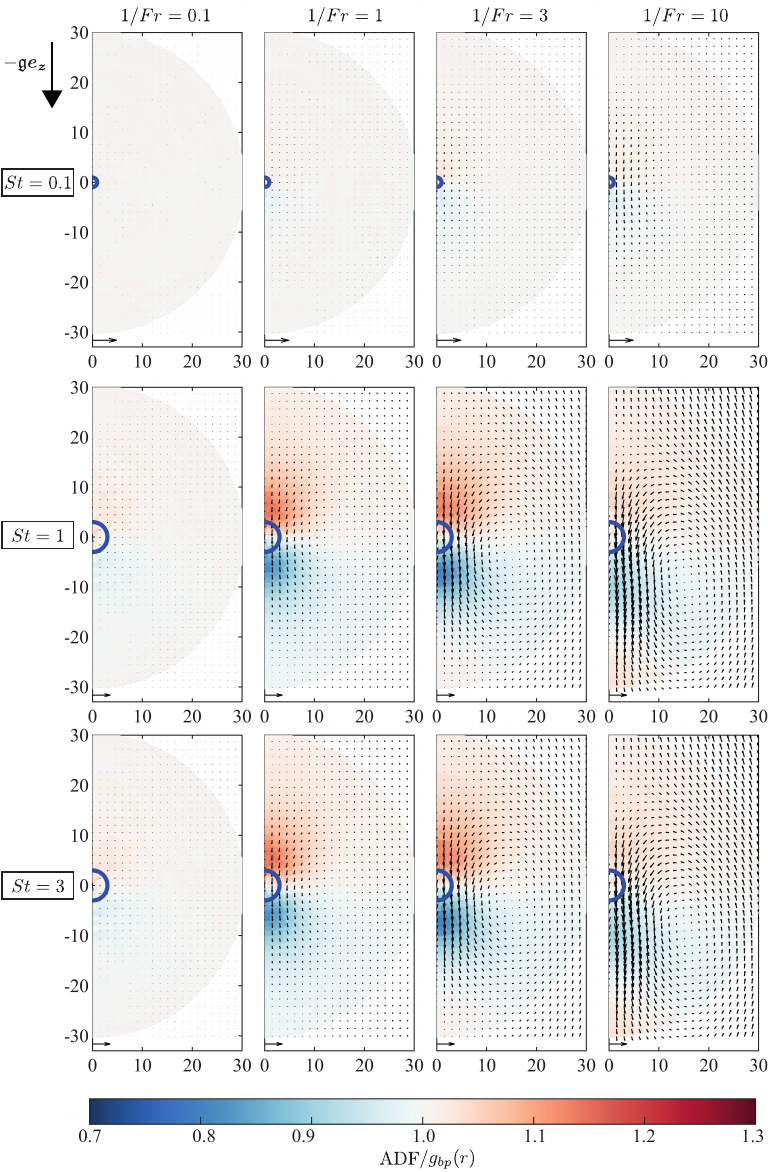}
    \caption{The angular distribution function at different distances and the azimuthally-averaged local flow field in a reference frame where the mean vertical flow velocity in the sampled region is zero. $\Rey_\lambda = 167$. The semicircle and the reference arrow on the bottom row indicate the bubble and $u_\eta$, respectively. The axis labels are in units of $\eta$.}
    \label{fig:ADF}
\end{figure}

As gravity introduces anisotropy in the vertical direction, we further quantitatively examine the bubble--particle spatial distribution by binning the RDF by the polar angle to obtain the angular distribution function (ADF) \citep{gualtieri_anisotropic_2009,ireland_effect_gavity_2016}, which is normalised by the RDF in figure \ref{fig:ADF}. ADF$/g_{bp}(r) \neq 1$ everywhere when gravity is included as expected \citep{ireland_effect_gavity_2016}. Intriguingly, particles preferentially sample the regions directly above bubbles, especially when $St/Fr$ becomes sufficiently large. To understand this, we consider figure \ref{fig:fluidVelPpos}(\textit{a}) which showed that bubbles and particles preferentially sample downflows, and examine the local flow structure by overlaying the ADF on top of the azimuthally-averaged flow field around the bubble in a reference frame where the mean vertical flow velocity in the sampled region is zero in figure \ref{fig:ADF}. The results show that there are vortices with their downward branches passing the bubble and the region where the flow converges horizontally corresponds to the locations with higher particle concentration, suggesting that the convergent flow influences the particle trajectories such that they are concentrated above bubbles. In the cases with appreciable anisotropic particle distributions (i.e. $St = 1,3$ and $1/Fr \geq 1$), the vertical component of the flow velocity vector at the bubble position as displayed in figure \ref{fig:ADF} amounts to 60-80\% of $\langle u_z \rangle_b$ and decreases very slightly as $1/Fr$ increases. Despite the local flow field contributing less to $\langle u_z \rangle_b$ with increasing $1/Fr$, the particle distribution is most anisotropic at intermediate $1/Fr$ since the particles settle faster at large $1/Fr$, meaning they will have less time to interact with the background turbulence such that they will be more isotropically distributed. For the tested parameters, this anisotropy increases with $St$. Nonetheless, we expect that it is strongest for some finite $St$ and weakens as $St$ becomes very large when the particles no longer respond to the background flow. Overall, this means 
that collisions will occur more frequently on top of a bubble at intermediate $1/Fr$ and $St$ even without considering the azimuthal variation of the collision velocity due to gravity.

\subsection{Collision velocity} \label{sec::collisionVelocity_G}
\begin{figure}
    \centering
    \includegraphics{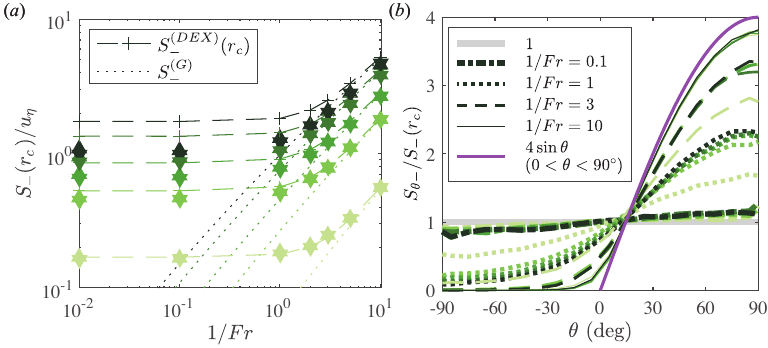}
    \caption{(\textit{a}) $S_-(r_c)$ at different $1/Fr$. The model predictions are displayed only for $\Rey_\lambda = 167$. The colour scheme and symbols follow figure \ref{fig:collK_G}(\textit{a}). (\textit{b}) The nondimensionalised distribution of $S_{\theta-}$ along the polar angle $\theta$ for $\Rey_\lambda = 167$. The colour scheme follows figure \ref{fig:collK_G}(\textit{a}).}
    \label{fig:colVel_G}
\end{figure}

Besides the bubble--particle distribution, the collision kernel is determined by the effective radial approach velocity at collision distance $S_-(r_c)$ according to (\ref{eq:CollisionKernelKinematic_4pi_G}). Figure \ref{fig:colVel_G}(\textit{a}) shows $S_-(r_c)$ as a function of $1/Fr$. $S_-$ increases significantly with $1/Fr$ when $1/Fr > 1$ for all $St$ and almost recovers the relative settling limit $S_-^{(G)}$ at $1/Fr \sim 10$ --- a point which we will return to. At low $1/Fr$ where turbulence dominates, $S_-$ differs across $St$ because of the different collision mechanisms involved. At small $St$, collisions occur due to local fluid shear and the local turnstile effect as introduced in \S\ref{sec::intro_G}. As $St$ increases, local effects become less important since the bubbles and particles interact with larger and more energetic eddies. Due to the different physical properties of bubbles and particles, the interaction is different between bubbles and particles which contributes to $S_-$, i.e. the `nonlocal turnstile effect'. The local shear and local turnstile effects, as well as relative settling effect of gravity, are properly represented in the extended \citet{dodin_collision_2002} prediction $S_-^{(DEX)}$, which explains its excellent agreement with the data 
throughout the entire $1/Fr$ range at $St = 0.1$. However, it increasingly overpredicts $S_-$ at larger $St$ as it does not model any nonlocal effects. These nonlocal effects do not directly increase $S_-$ by improving the alignment between the relative velocity vector and the separation vector, in contrast to the local turnstile effect. Considering only local effects at higher $St$ regimes can therefore lead to the overprediction. At higher $1/Fr$, the overprediction decreases since the contribution of the turbulence collision mechanisms decreases and relative settling plays an increasingly dominant role. 
Another approach to quantify the gravity effect is to consider the angular distribution of $S_-(r_c)$ with respect to gravity, where $S_{-}(r_c)$ is binned using the polar angle $\theta$ around the bubble to obtain $S_{\theta-}$. This is plotted in figure \ref{fig:colVel_G}(\textit{b}) with $\theta = -90^\circ$ corresponding to the direction of gravity. At the no-gravity ($1/Fr = 0$) limit, there is no preferred direction and accordingly $S_{\theta-}$ is uniformly distributed across all $\theta$ (grey line). As $1/Fr$ increases, more and more collisions occur on the top half of a bubble where the approach velocity is enhanced due to gravitational settling and vice-versa for the lower half. Consequently, the top hemisphere ($\theta>0^\circ$) contributes more to $S_-(r_c)$ than the bottom one. This anisotropy is still very weak at $1/Fr = 0.1$ and becomes significant at $1/Fr = 1$, even though at this value $S_-(r_c)$ still has not increased significantly from the turbulence-dominated limit as shown in figure \ref{fig:colVel_G}(\textit{a}). At $1/Fr = 10$, only a small fraction of the collisions occur in the bottom hemisphere due to turbulence, and the distribution is close to the relative settling limit (purple line) where collisions exclusively occur on the top hemisphere. 
Furthermore, between $1\leq 1/Fr< 10$, $S_{\theta-}$ is not sensitive to $St$ for $St \geq 0.5$, suggesting the transition from turbulence to relative settling for the collision velocity is similar at different values of $St$.

\begin{figure}
    \centering
    \includegraphics[]{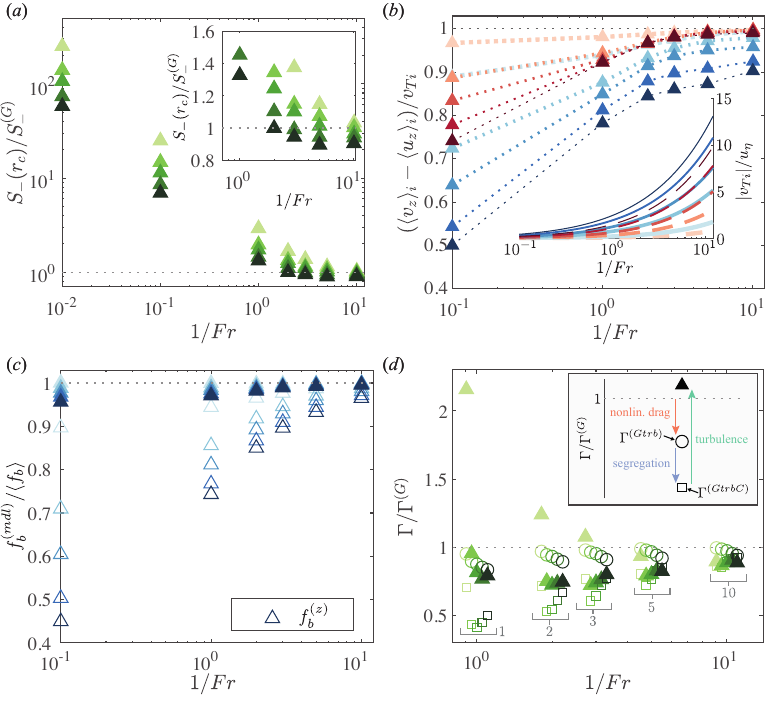}
    \caption{(\textit{a}) The ratio of $S_-(r_c)$ to the relative settling case $S_-^{(G)}$ at $\Rey_\lambda = 167$. Inset shows an enlarged version of the region where $S_-(r_c) < S_-^{(G)}$. (\textit{b}) The ratio of the average bubble and particle vertical slip velocity to the terminal velocity in still fluid accounting for nonlinear drag $v_{Ti}$, whose magnitude is shown in the inset (solid line for bubbles and dashed lines for particles), at $\Rey_\lambda = 167$. The lines in the main figure are guides for the eye. (\textit{c}) The ratio of the modelled drag correction to the measured drag correction at $\Rey_\lambda = 167$. The open symbols indicate the drag correction based solely on the vertical slip velocity. (\textit{d}) The collision kernel normalised by the relative settling case in still fluid at $\Rey_\lambda = 167$. Also shown are the relative settling collision kernel in turbulence $\Gamma^{(Gtrb)}$ and the effect of segregation. The cases with different $St$ have been laterally displaced for clarity and the associated labels indicate the corresponding $1/Fr$. The symbols and the colour scheme in all the panels follow figures \ref{fig:collK_G}(\textit{a}) and \ref{fig:fluidVelPpos}(\textit{a}).}
    \label{fig:reducedriseBubble}
\end{figure}

Similar to our analysis of the collision kernel, we normalise $S_-(r_c)$ with the relative settling case in still fluid $S_-^{(G)} = (|v_{Tb}| + |v_{Tp}|)/4$, where $v_{Ti}$ are the terminal velocities obtained by balancing the drag and buoyancy terms in (\ref{eq:PointParticleEoM_G}) taking $\boldsymbol{u} = \mathbf{0}$, and plot the result in figure \ref{fig:reducedriseBubble}(\textit{a}). Although we only show the data for the $Re_\lambda = 167$ case, the results are not sensitive to $Re_\lambda$ so the following discussion also applies to $Re_\lambda = 69$. Generally, $S_-(r_c)/S_-^{(G)}$ decreases as $1/Fr$ increases, reflecting the growing contribution of gravity. Upon closer inspection, we observe that $S_-(r_c)$ for $St \geq 0.5$ dips below $S_-^{(G)}$, instead of simply approaching $S_-^{(G)}$ as is the case for $St = 0.1$. Note that this is not due to preferential sampling as the fluid velocity sampled by the colliding bubbles and particles are almost the same as already shown in figure \ref{fig:fluidVelPpos}(\textit{b}).
To investigate why $S_- < S_-^{(G)}$, we compare the mean vertical slip velocity of the bubbles and particles $\langle v_z \rangle_i - \langle u_z \rangle_i$ with the theoretical terminal velocity in still fluid $v_{Ti}$, on which $S_-^{(G)}$ in figure \ref{fig:reducedriseBubble}(\textit{b}) is based on. For $St = 0.1$, $(\langle v_z \rangle_i - \langle u_z \rangle_i)/v_{Ti}$ is very close to 1 throughout, which is consistent with the fact that $S_- \geq S_-^{(G)}$. For higher $St$, however, the mean vertical slip velocity in turbulence is smaller than $v_{Ti}$. This effect is generally stronger for bubbles, more pronounced at high $St$ and overall the ratio $(\langle v_z \rangle_i - \langle u_z \rangle_i)/v_{Ti}$ trends back up towards 1 as buoyancy becomes stronger. 
To understand its origin, we focus on the bubbles and consider the fact that turbulence induces additional horizontal slip velocities. As $Re_i$ depends on magnitude of the total slip velocity, the horizontal slip increases the value of $Re_i$ and the drag force, which can explain the slowdown \citep{ruth_effect_2021}. To test this hypothesis, we model the slip velocity vector as $(\boldsymbol{v_b}-\boldsymbol{u})^{(mdl)} = (v_{bx} - u_{bx})'\boldsymbol{e_x} + (v_{bx} - u_{bx})'\boldsymbol{e_y} + (\langle v_z\rangle_b - \langle u_z\rangle_b)\boldsymbol{e_z}$, where $(v_{bx} - u_{bx})'$ is the measured r.m.s. of the horizontal bubble slip velocity and $\boldsymbol{e_{x,y,z}}$ are unit vectors along $x$-, $y$- and the vertical directions, respectively. Based on the magnitude of the slip velocity vector, we then calculate the resulting $Re_i$ and the corresponding drag correction term $f_b^{(mdl)}$. Figure \ref{fig:reducedriseBubble}(\textit{c}) shows that $f_b^{(mdl)}$ is indeed close to the measured mean drag correction $\langle f_b \rangle$. 
For reference, we also include the drag correction based on the vertical slip velocity only $(\boldsymbol{v_b}-\boldsymbol{u})^{(z)}=(\langle v_z\rangle_b - \langle u_z\rangle_b)\boldsymbol{e_z}$ as open symbols. 
As expected, $f_b^{(z)} < f_b^{(mdl)}$ and the difference reduces at larger $1/Fr$, consistent with the trend of $(\langle v_z \rangle_b - \langle u_z \rangle_b)/v_{Tb}$ in figure \ref{fig:reducedriseBubble}(\textit{b}). This is because the mean vertical slip velocity becomes larger relative to the horizontal slip velocities with increasing $1/Fr$, such that $|\langle v_z\rangle_b - \langle u_z\rangle_b|$ increasingly dominates $|\boldsymbol{v_b}-\boldsymbol{u}|$. We therefore conclude that the observed reduction in the rise velocity is indeed due to enhanced nonlinear drag caused by the horizontal slip velocity components in turbulence. 

Combining these insights, we re-examine the unexpected result that turbulence can reduce the collision rate (i.e. $\Gamma < \Gamma^{(G)}$), as seen in figure \ref{fig:collK_G}(\textit{a}). Our analysis shows that this can be attributed to bubble--particle spatial segregation and the reduced bubble slip velocity in turbulence. To evaluate their respective importance, we again show $\Gamma/\Gamma^{(G)}$ in figure \ref{fig:reducedriseBubble}(\textit{d}), this time focusing only on the parameter range where $\Gamma < \Gamma^{(G)}$. To quantify the effect of the nonlinear drag, we plot the relative settling collision kernel in turbulence $\Gamma^{(Gtrb)} = \pi r_c^2 [\langle v_{z} \rangle_b - \langle u_{z} \rangle_b - (\langle v_{z} \rangle_p - \langle u_{z} \rangle_p)]$, i.e. based on the measured vertical slip velocities of bubbles and particles. $\Gamma^{(Gtrb)}/\Gamma^{(G)}\leq 1$ as the nonlinear drag reduces the bubble slip velocity. As $1/Fr$ increases, $\Gamma^{(Gtrb)}/\Gamma^{(G)}$ increases since the magnitude of the slip velocity is increasingly dominated by the vertical component. Whereas, $\Gamma^{(Gtrb)}/\Gamma^{(G)}$ decreases for increasing $St$ as the effect of nonlinear drag becomes more pronounced. Since bubbles and particles segregate in turbulence, we also show $\Gamma^{(GtrbC)} = \Gamma^{(Gtrb)}\cdot g(r_c)$ in the figure. The difference between $\Gamma^{(GtrbC)}$ and $\Gamma^{(Gtrb)}$ decreases with increasing $1/Fr$, reflecting the fact that gravity reduces segregation. Finally, the difference between $\Gamma^{(GtrbC)}$ and $\Gamma$ indicates the enhancement of the collision velocity due to turbulence mechanisms. This is larger at lower $1/Fr$ and by $1/Fr = 10$, $\Gamma \approx \Gamma^{(GtrbC)}$.

{\subsection{Effect of the particle density}    \label{sec::rhopInf_G}}

\begin{figure}
    \centering
    \includegraphics[]{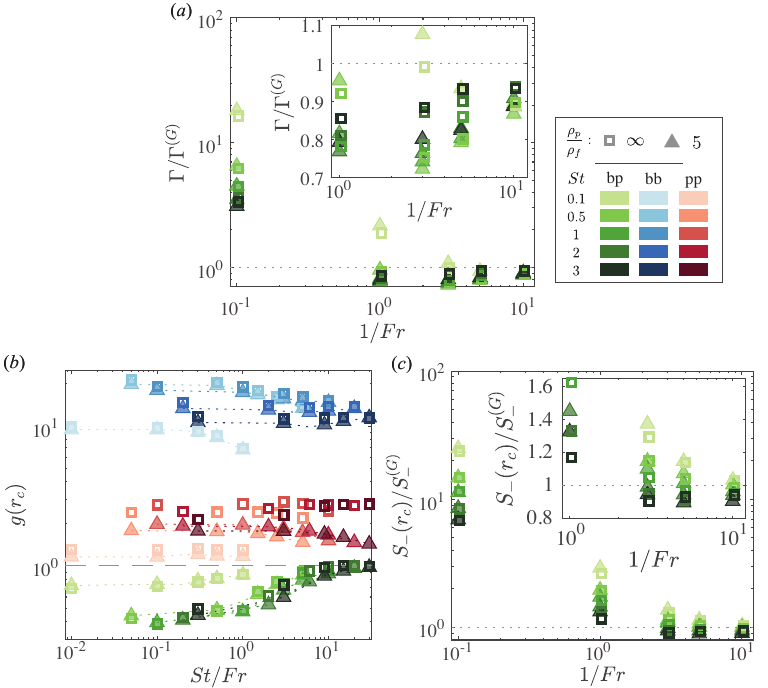}
    \caption{{(\textit{a}) The collision kernel $\Gamma$ normalised by the relative settling limit, (\textit{b}) the radial distribution function $g(r_c)$, and (\textit{c}) $S_-(r_c)$ normalised by the relative settling limit for both $\rho_p/\rho_b = 5$ (filled triangles) and $\rho_p/\rho_b =\infty$ (open squares) at $Re_\lambda = 167$. The insets in (\textit{a}) and (\textit{c}) zooms in to the $1 \lesssim 1/Fr \lesssim 10$ region.}}
    \label{fig:rhopInf_G}
\end{figure}

{In the previous sections, we fixed $\rho_p/\rho_f = 5$ and vary $St$ and $1/Fr$. Nevertheless, $\rho_p/\rho_f$, which sets $r_p$ hence $r_c$, can have an explicit effect on the particle dynamics beyond determining the collision distance according to (\ref{eq:PointParticleEoM_G}). We therefore additionally simulate $\rho_p/\rho_f = \infty$ with $1/Fr = 0.1,1,3,5$, and 10 for all the tested values of $St$ at $Re_\lambda = 167$, while retaining the same $r_c$ as the $\rho_p/\rho_f=5$ case. Figure \ref{fig:rhopInf_G} plots the collision kernel $\Gamma$, the RDF $g(r_c)$, and the effective radial approach velocity $S_-(r_c)$, in which $\Gamma$ and $S_-(r_c)$ have been normalised by the relative settling limit to account for the different particle settling velocities. These data show that for the bubble--particle case neither the overall collision kernel $\Gamma/\Gamma^{(G)}$, nor the  factors $g_{bp}(r_c)$, and $S_-(r_c)/S_-^{(G)}$ individually are sensitive to $\rho_p$. This is in contrast to the particle--particle case where $g_{pp}(r_c)$ remains constant or even increases slightly with $1/Fr$, which is shown in figure \ref{fig:rhopInf_G}(\textit{b}) and is also observed by \citet{woittiez_combined_2009}. Despite $g_{pp}(r_c)$ remaining above 1, $g_{bp}(r_c)$ approaches unity similar to the $\rho_p/\rho_f = 5$ case, which is consistent with our interpretation of figure \ref{fig:RDF_G}(\textit{a}) in \S\ref{sec::bpDistribution_G} that the trend in $g_{bp}(r_c)$ with $1/Fr$ is not principally due to a change in bubble or particle clustering strength. We furthermore note that even for $\rho_p/\rho_f = \infty$, $S_-(r_c) < S_-^{(G)}$ and $\Gamma < \Gamma^{(G)}$ at intermediate values of $1/Fr$. This demonstrates that the reduction of the bubble--particle collision rate below the pure gravity case by turbulence is not specific to particles with $\rho_p/\rho_f = 5$.}

\begin{figure}
    \centering
    \includegraphics[]{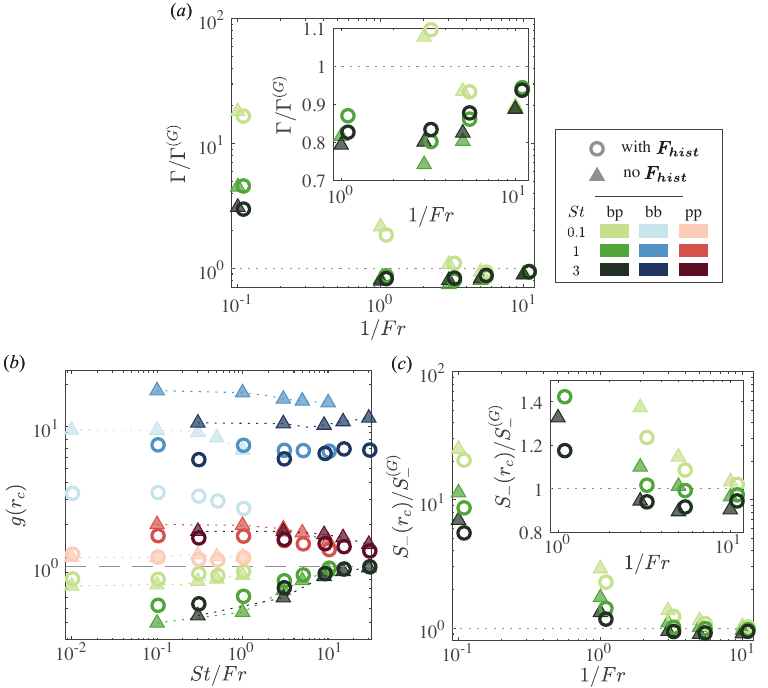}
    \caption{{(\textit{a}) The collision kernel $\Gamma$ normalised by the relative settling limit, (\textit{b}) the radial distribution function $g(r_c)$, and (\textit{c}) $S_-(r_c)$ normalised by the relative settling limit for the cases with and without the history force $\boldsymbol{F_{hist}}$ at $Re_\lambda = 167$. The insets in (\textit{a}) and (\textit{c}) zooms in to the $1 \lesssim 1/Fr \lesssim 10$ region.}}
    \label{fig:histForc_G}
\end{figure}

{\subsection{Effect of the history force}   \label{sec::historyForce_G}}
{In addition to testing $\rho_p/\rho_f = \infty$, we examine the effect of the history force by including (\ref{eq:bassetHistoryForce}) in (\ref{eq:PointParticleEoM_G}) and simulating $St = 0.1,1,$ and 3 for $1/Fr = 0.1,1,3,5,10$ at $Re_\lambda = 167$. The measured $\Gamma/\Gamma^{(G)}$, $g(r_c)$, and $S_-(r_c)/S_-^{(G)}$ are plotted in figure \ref{fig:histForc_G}. These results show that the history force does not qualitatively change the bubble--particle collision statistics, insofar as the general trends with $1/Fr$ remains the same, the reduction of $\Gamma$ and $S_-(r_c)$ below the relative settling limit is still observed in figures \ref{fig:histForc_G}(\textit{a}) and \ref{fig:histForc_G}(\textit{c}), and $g_{bp}(r_c)$ still lies below 1 at low to intermediate values of $1/Fr$ as shown by figure \ref{fig:histForc_G}(\textit{b}). Quantitatively, the history force does not significantly affect the collision kernel, since it tends to reduce $S_-(r_c)$ and increase the value of $g_{bp}(r_c)$. This increase in the value of $g_{bp}(r_c)$ towards 1 corresponds to reduced bubble--particle segregation. It is hence consistent with the lower values of $g_{bb}(r_c)$ and $g_{pp}(r_c)$ observed when the history force is included in figure \ref{fig:histForc_G}(\textit{b}), which indicate weaker bubble and particle clusters and have already been reported in the literature \citep{daitche_role_2015,olivieri_effect_2014}. In summary, our results show that the history force does not play a role in determining the qualitative trends of the bubble--particle collisions statistics and only a minor one for the value of the collision kernel.}

\section{Discussion and Conclusion}  \label{sec::conclusion_G}
We studied the effect of gravity on bubble--particle collisions in HIT by conducting direct numerical simulations using the point-particle approach {\citep{maxey_equation_1983,gatignol_faxen_1983,auton_lift_1987,auton_force_1988}}. From our simulations, we find that gravity is negligible and turbulence mechanisms determine the collision kernel up to $1/Fr = 0.1$. For instance, bubbles and particles form individual clusters and segregate in this regime. When $1 \lesssim 1/Fr < 10$, gravity noticeably influence the collision kernel even though the effects of turbulence are still significant. Gravity increases the collision kernel by reducing the extent of segregation and increasing the collision velocity since bubbles rise and particles settle. 
Notably, the collision kernel dips below the still fluid case when $1/Fr \gtrsim 5$ for $St = 0.1$ and $1/Fr \gtrsim 1$ for $St \geq 0.5$, meaning turbulence reduces the collision rate in this regime. This is due to preferential sampling and additionally for $St \geq 0.5$ a reduction in the bubble slip velocity due to nonlinear drag. Since both effects weaken as $1/Fr$ increases, turbulence effects become negligible and the collision kernel is well approximated by the still fluid case when $1/Fr \geq 10$. {For our tested parameters, additional simulations reveal that the observed trends of the bubble--particle collision statistics are not sensitive to particle density and the history force. Quantitatively, the history force tends to reduce bubble--particle segregation and the effective bubble--particle radial approach velocity. These two phenomena have opposite effects on the collision kernel, such that the net effect of the history force on the collision kernel is weak. } 
Although the existing bubble--particle collision models qualitatively capture the increasing trend in the collision velocity with $1/Fr$ over the tested range, none of them achieve quantitative agreement with the simulation data in terms of the collision kernel {even when the predictions are appropriately compared to simulations without the history force}. We extended the model by \citet{dodin_collision_2002} to the bubble--particle case and found excellent agreement with our simulations over all $1/Fr$ for small $St$ if segregation is accounted for {and when the history force is neglected}. Note that none of these models capture $S_- < S_-^{(G)}$ which occurs at higher $St$ due to the effect of nonlinear drag. This leads to a discrepancy of up to 20\% in $S_-$ in the tested regime.

Throughout this paper, we have used both $1/Fr$ and $St/Fr$ in order to parameterise the turbulence-to-gravity transition and the gravity-dominated limit. As discussed, gravity starts to play a noticeable role when $1/Fr \gtrsim 1$. To understand why $1/Fr$ is the relevant parameter, first take $\boldsymbol{\Delta v}$ as a proxy for $S_-$ and consider the following heuristic argument: at small $St$, the bubble--particle relative velocity $\boldsymbol{\Delta v}$ is given by the shear mechanism which is $\propto \sqrt{St}$, the local turnstile mechanism which is $\propto St$, and the relative settling contribution which is $\propto St/Fr$ (see (\ref{eq:DeltaV_G})). Assuming the shear mechanism is weaker than the local turnstile mechanism, $\boldsymbol{\Delta v} \propto St + St/Fr$. As $(\boldsymbol{\Delta v})/(\boldsymbol{\Delta v}|_{1/Fr=0}) \propto 1 + 1/Fr$, we consider $1/Fr$ when examining the turbulence-to-gravity transition. On the other hand, $St/Fr$ is the appropriate nondimensional number for the gravity-dominant limit as relative settling is governed by $St/Fr$ \citep{good_settling_2014}.

Multiple questions still remain to be addressed. Our simulations are conducted using the point-particle approximation, {which helps to provide a first-order appreciation of bubble--particle collisions and enables one-to-one comparison with the geometric collision models. As a trade-off for its relative simplicity, this approach limits access to the large-$St$ regime and inherently} neglects finite-size effects. {For the largest values of $St$ tested $r_b/\eta \sim 5$, which is beyond the commonly adopted bounds of the point-particle approximation \citep{homann_finite-size_2010}. Therefore, our results at these values should be interpreted with caution as finite size effects may be relevant.} Performing interface-resolved simulation using for example the immersed boundary method \citep{verzicco_immersed_2023} can provide additional insights as the bubble distorts the local flow field and can further modify the bubble--particle distribution \citep{tiedemann_collision_2023,jiang_how_2024}{, though the extra complexity would make it harder to distinguish whether the trends in the collision statistics are due to the locally deformed flow field or due to the background turbulence.} Furthermore, we considered the {more straightforward} case of bubbles and particles with the same $St$. While this has provided much understanding into the collision process, unravelling the collision mechanisms for bubbles and particles with different $St$ would prove invaluable for real life applications, where particles in flotation cells are usually far from monodispersed. In terms of modelling, a theoretical model of segregation will greatly enhance the predictive capabilities of the existing models.

\section*{Acknowledgements}
We thank Giulia Piumini, Duco van Buuren, Kevin Zhong, and Giuseppe Vacca for helpful discussions.

\section*{Funding}
This project has received funding from the European Research Council (ERC) under the European Union's Horizon 2020 research and innovation programme (grant agreement No. 950111, BU-PACT). This work was carried out on the Dutch national e-infrastructure with the support of SURF Cooperative. We acknowledge the EuroHPC Joint Undertaking for awarding the project EHPC-REG-2023R03-178 access to the EuroHPC supercomputer Discoverer, hosted by Sofia Tech Park (Bulgaria).
 
\section*{Declaration of interests}
The authors report no conflict of interest.

\section*{Data availability statement}
All data supporting this study are available from the authors upon request.

\appendix
\section{Consistent inclusion of gravity following \citet{abrahamson_collision_1975}}    \label{sec::AbrahamsonincludeGravity}
The original expression of the collision kernel including gravity as given in \citet{abrahamson_collision_1975} has an integration error which has been partly resolved by \citet{kostoglou_generalized_2020} using a comparable expression. However, the equivalence between the expressions in \citet{abrahamson_collision_1975} and \citet{kostoglou_generalized_2020} was not directly shown. We provide this additional detail in this appendix. Furthermore, we show the difference between (\ref{eq:AbrahamsonGravIntegrationFit}) and the best-fit expressions reported in figure 2 and (10) in \citet{kostoglou_generalized_2020}.

According to \citet{abrahamson_collision_1975}, the collision kernel between two types of particles in the presence of gravity is given by 
\begin{eqnarray}
    \Gamma_{12} &=& \pi r_c^2 v_c\nonumber\\ &=& \frac{\pi r_c^2}{(2\pi v_1'^2)^{3/2}(2\pi v_2'^2)^{3/2}}\multiint{6}\limits_{\mathrm{all\,space}} \sqrt{(v_{1x}-v_{2x})^2 + (v_{1y}-v_{2y})^2 + (v_{1z}-v_{2z})^2}\nonumber\\
    && \times\exp{\bigg[\sum_{i = 1}^{2}-\frac{v_{ix}^2 + v_{iy}^2 + (v_{iz} - v_{Ti})^2}{2v_i'^2} \bigg]} \mathrm{d}v_{1x}\mathrm{d}v_{1y}\mathrm{d}v_{1z}\mathrm{d}v_{2x}\mathrm{d}v_{2y}\mathrm{d}v_{2z},
\end{eqnarray}
where the numerical subscripts $1,2$ denote the colliding particles, and the alphabetical subscripts $x,y,z$ denote the corresponding velocity components. Changing variables such that $v_{iN} = v_{iz} - v_{Ti}$ for $i = 1,2$ and further taking
\begin{subeqnarray}
    v_{xm} = \frac{v_{1x} + v_{2x}}{2},
   \quad v_{ym} = \frac{v_{1y} + v_{2y}}{2}, \quad 
    v_{zm} = \frac{v_{1N} + v_{2N}}{2}\\
    v_{xd} = v_{1x} - v_{2x},
   \quad v_{yd} = v_{1y} - v_{2y}, \quad 
    v_{zd} = v_{1N} - v_{2N}
\end{subeqnarray}
yields
\begin{eqnarray}
    v_c &=& \frac{1}{(2\pi v_1'^2)^{3/2}(2\pi v_2'^2)^{3/2}}\iiint\limits_{\mathrm{all\,space}} \sqrt{v_{xd}^2 + v_{yd}^2 + [v_{zd} + (v_{T1} - v_{T2})]^2}\nonumber\\
    && \times I_x \cdot I_y \cdot I_z\cdot \mathrm{d}v_{xd}\mathrm{d}v_{yd}\mathrm{d}v_{zd},
\end{eqnarray}
where
\begin{equation}
    I_j = \int_{-\infty}^{+\infty} \exp\bigg[-\frac{v_1'^2 + v_2'^2}{2v_1'^2v_2'^2}\bigg( v_{jm} - \frac{v_2'^2 - v_1'^2}{2(v_1'^2 + v_2'^2)} v_{jd}\bigg)^2 \bigg] \exp\bigg[-\frac{v_{jd}^2}{2(v_1'^2 + v_2'^2)} \bigg] \mathrm{d}v_{jm}
\end{equation}
and $j = x,y,z$. Using $\int_{-\infty}^{+\infty}\exp[-a(x-b)^2]\mathrm{d}z = \sqrt{\pi/a}$ where $a,b$ are constants to evaluate $I_x$, $I_y$ and $I_z$,
\begin{eqnarray}    \label{eq:KostoglouEquivalentIntegral}
    v_c &=& \frac{1}{[2\pi(v_1'^2 + v_2'^2)^{3/2}]}\iiint\limits_{\mathrm{all\,space}} \sqrt{v_{xd}^2 + v_{yd}^2 + [v_{zd} + (v_{T1} - v_{T2})]^2}\nonumber\\ && \times \exp\bigg[ -\frac{v_{xd}^2 + v_{yd}^2 + v_{zd}^2}{2(v_1'^2 + v_2'^2)} \bigg] \mathrm{d}v_{xd}\mathrm{d}v_{yd}\mathrm{d}v_{zd},
\end{eqnarray}
which is the same as (8) in \citet{kostoglou_generalized_2020} with the following mapping:
\begin{equation*}
    \sqrt{v_1'^2 + v_2'^2} \to V,
   \quad v_{T2} - v_{T1} \to U_B, \quad 
    v_c \to U_T, \quad \frac{v_{T2} - v_{T1}}{v_1'^2 + v_2'^2} \to \alpha, \quad \frac{v_c}{\sqrt{v_1'^2 + v_2'^2}} \to \Psi.
\end{equation*}

\begin{figure}
    \centering
    \includegraphics[]{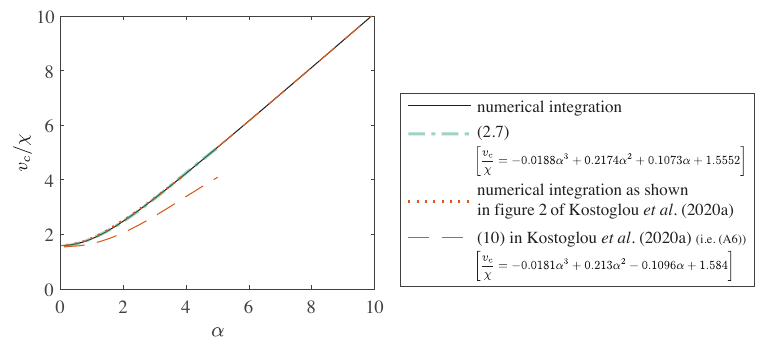}
    \caption{Plot of $v_c/\chi$ as a function of $\alpha$. Note that (\ref{eq:AbrahamsonGravIntegrationFit}) differs from (\ref{eq:KostoglouBestFit}) as the fitting coefficients are different for $0.1\leq \alpha \leq 5$. The respective fits in this range are displayed in the legend.}
    \label{fig::abrahamsonIntegration_KostoglouFit}
\end{figure}

\citet{kostoglou_generalized_2020} numerically integrated (\ref{eq:KostoglouEquivalentIntegral}), showed a best-fit result in their figure 2 and stated in their (10) that the corresponding equation is
\begin{equation}    \label{eq:KostoglouBestFit}
\frac{v_c}{\chi} = 
\left\{
\begin{aligned}
1.6,     \quad&\text{for $\alpha < 0.1$}\\
-0.0181\alpha^3+0.213\alpha^2-0.1096\alpha+1.584,\quad&\text{for $0.1\leq \alpha\leq 5$}\\
\alpha + (1/\alpha),\quad&\text{for $\alpha>5$}
\end{aligned}
\right.
\end{equation}
where $\alpha = (v_{T2}-v_{T1})/\chi$ and $\chi = \sqrt{v'^{2}_1 + v'^{2}_2}$.
Both the curve shown in their figure 2 and (\ref{eq:KostoglouBestFit}) are plotted in figure \ref{fig::abrahamsonIntegration_KostoglouFit}, which clearly indicates their difference. We additionally include our best-fit expression (\ref{eq:AbrahamsonGravIntegrationFit}) which shows excellent agreement with the numerical integration result as displayed in figure 2 of \citet{kostoglou_generalized_2020}.

\section{Extension of the \citet{dodin_collision_2002} model to particles with different densities}    \label{sec::extDodinElperin_bpG}
The \citet{dodin_collision_2002} model is applicable for infinitely heavy inertial particles with a uniform spatial distribution in HIT. Here we extend it to particles with different densities by incorporating the density ratio $\beta_i = 2(\rho_f - \rho_i)/(\rho_f + 2\rho_i)$.

We start from (\ref{eq:PointParticleEoM_G}) in the small $St$ limit without lift \citep{fouxon_distribution_2012}
\begin{equation}
    \boldsymbol{v_i} = \boldsymbol{u} + \frac{\beta_i\tau_i}{f_i}\boldsymbol{\psi} + \frac{\beta_i\tau_i}{f_i}\mathfrak{g}\boldsymbol{e_z}
\end{equation}
where $\boldsymbol{\psi} = D\boldsymbol{u}/Dt$. Consider the case where two particles located at positions $A$ and $B$ are at collision distance such that their separation is small. $\boldsymbol{\psi}|_A\approx\boldsymbol{\psi}|_B$ and
\begin{equation}    \label{eq:DeltaV_G}
   \boldsymbol{\Delta v} = \boldsymbol{v_2} - \boldsymbol{v_1} = \underbrace{(\boldsymbol{u}|_B - \boldsymbol{u}|_A) + \bigg(\frac{\beta_2\tau_2}{f_2} - \frac{\beta_1\tau_1}{f_1}\bigg)\boldsymbol{\psi}|_A}_{\boldsymbol{\Delta v_f}} + \underbrace{\bigg(\frac{\beta_2\tau_2}{f_2} - \frac{\beta_1\tau_1}{f_1}\bigg)\mathfrak{g}\boldsymbol{e_z}}_{\boldsymbol{\Delta v_g}},
\end{equation}
in which $\boldsymbol{\Delta v}$ has been split into fluid $\boldsymbol{\Delta v_f}$ and gravity contributions $\boldsymbol{\Delta v_g}$. Projecting (\ref{eq:DeltaV_G}) along the separation vector, of which $\boldsymbol{e_r}$ is the corresponding unit vector, gives
\begin{equation}
    \Delta v_r(\theta) = (\boldsymbol{\Delta v_f} + \boldsymbol{\Delta v_g})\cdot \boldsymbol{e_r} \coloneq \xi + h(\theta) = \xi + \bigg(\frac{\beta_2\tau_2}{f_2} - \frac{\beta_1\tau_1}{f_1}\bigg)\mathfrak{g}\cos\theta.
\end{equation}

As the background flow is HIT, $\xi$ is isotropic so it is assumed to be normally distributed. Furthermore, since $h(\theta)$ has a sign ambiguity originating from the possibility of labelling either particle as number 1, we introduce 
\begin{equation}
    h_+(\theta) = \bigg|\frac{\beta_2\tau_2}{f_2} - \frac{\beta_1\tau_1}{f_1}\bigg|\mathfrak{g}\cos\theta
\end{equation}
and write
\begin{eqnarray}
    \langle|\Delta v_r(\theta)|\rangle &=& \int_{-\infty}^{+\infty} \frac{|\alpha|}{\sigma\sqrt{2\pi}}\exp{\bigg(-\frac{(\alpha - h_+)^2}{2\sigma^2}\bigg)}d\alpha\nonumber\\
    &=& \sigma\sqrt{2}\kappa\erf\kappa + \sigma\sqrt{\frac{2}{\pi}}\exp{(-\kappa^2)},
\end{eqnarray}
where $\langle \cdot \rangle$ denotes averaging, $\kappa = h_+/(\sigma\sqrt{2})$ and $\sigma = \sqrt{\langle\xi^2\rangle} =\sigma_{\Delta vr}^{(DEX)}$ is the r.m.s. of $\xi$ (we drop the subscript and superscript of $\sigma_{\Delta vr}^{(DEX)}$ in this appendix to reduce clutter). For $\sigma$, we use the isotropy of the background flow and place particle 2 along the $x$-axis at position C. Similar to before, we consider particles at collision distance so $\boldsymbol{\psi}|_A \approx \boldsymbol{\psi}|_C$, meaning
\begin{eqnarray}
    \sigma^2 = \langle\xi^2\rangle &=& \langle(\boldsymbol{\Delta v_f}\cdot\boldsymbol{e_x})^2\rangle\nonumber\\
    &=& \langle(u_x|_C - u_x|_A)^2 \rangle + \bigg(\frac{\beta_2\tau_2}{\langle f_2 \rangle} - \frac{\beta_1\tau_1}{\langle f_1 \rangle}\bigg)^2\langle \psi_x^2|_A \rangle\nonumber\\
    && + 2\bigg(\frac{\beta_2\tau_2}{\langle f_2 \rangle} - \frac{\beta_1\tau_1}{\langle f_1 \rangle}\bigg)\langle(u_x|_C - u_x|_A)\psi_x|_A \rangle,
\end{eqnarray}
where $\psi_x = Du_x/Dt$. Additionally, as $\langle(u_x|_C - u_x|_A)\psi_x|_A\rangle \approx \langle{(u_x|_C - u_x|_A)\psi_x|_C}\rangle = - \langle{(u_x|_C - u_x|_A)\psi_x|_A}\rangle$ due to homogeneity, $\langle{(u_x|_C - u_x|_A)\psi_x|_A} \rangle\approx 0$. Therefore,
\begin{equation}
    \sigma^2 = r_c^2\bigg\langle{\bigg(\frac{\partial u_x}{\partial x}\bigg)^2}\bigg\rangle + \bigg(\frac{\beta_2\tau_2}{\langle f_2\rangle} - \frac{\beta_1\tau_1}{\langle f_1\rangle}\bigg)^2 \bigg\langle{\bigg(\frac{Du_x}{Dt}\bigg)^2}\bigg\rangle \coloneq r_c^2\gamma^2 + \bigg(\frac{\beta_2\tau_2}{\langle f_2\rangle} - \frac{\beta_1\tau_1}{\langle f_1\rangle}\bigg)^2\lambda^2,
\end{equation}
where $\gamma^2 = \langle (\partial u_x/\partial x)^2\rangle = \varepsilon/(15\nu)$ and $\lambda^2 = \langle(Du_x/Dt)^2\rangle = 1.3\varepsilon^{3/2}\nu^{-1/2}$.

Finally, note that in the case where $g(r_c)=1$,
\begin{equation}    \label{eq:collKernel_G_DE_Appendix}
    \Gamma_{12} = \frac{1}{2} \int_{0}^{\pi} \langle{|\Delta v_r(\theta)|}\rangle(2\pi r_c \sin\theta) r_c d\theta = \pi r_c^2 \int_{0}^{\pi} \langle{|\Delta v_r(\theta)|}\rangle\sin\theta d\theta,
\end{equation}
where the factor $1/2$ in the first equality singles out the inward flux of particles across the collision sphere assuming flux balance. This equation can be further simplified using symmetry of $\langle|\Delta v_r(\theta)|\rangle$:
\begin{equation}
    \langle|\Delta v_r(\pi - \theta)|\rangle = \langle|\xi - h(\theta)|\rangle = \langle|-\xi + h(\theta)|\rangle = \langle|\Delta v_r(\theta)|\rangle,
\end{equation}
where the final equality is attributed to the symmetry of the distribution of $\xi$ about 0. (\ref{eq:collKernel_G_DE_Appendix}) then becomes
\begin{equation}
    \Gamma_{12} = 2\pi r_c^2 \int_{0}^{\pi/2} \langle{|\Delta v_r(\theta)|}\rangle\sin\theta d\theta.
\end{equation}
Performing the integration results in
\begin{equation}
    \Gamma_{12} = \sqrt{8\pi}r_c^2\sigma f(c),
\end{equation}
where
\begin{equation}
    f(c) = \frac{\sqrt{\pi}}{2}\bigg(c + \frac{1}{2c}\bigg)\erf c + \frac{\exp(-c^2)}{2}
\end{equation}
and
\begin{equation}
    c = \frac{|\beta_2\tau_2/\langle f_2\rangle - \beta_1\tau_1/\langle f_1\rangle|\mathfrak{g}}{\sqrt{2}\sigma}.
\end{equation}

\section{Domain size effects}   \label{sec::domainSize_G}
\begin{figure}
    \centering
    \includegraphics{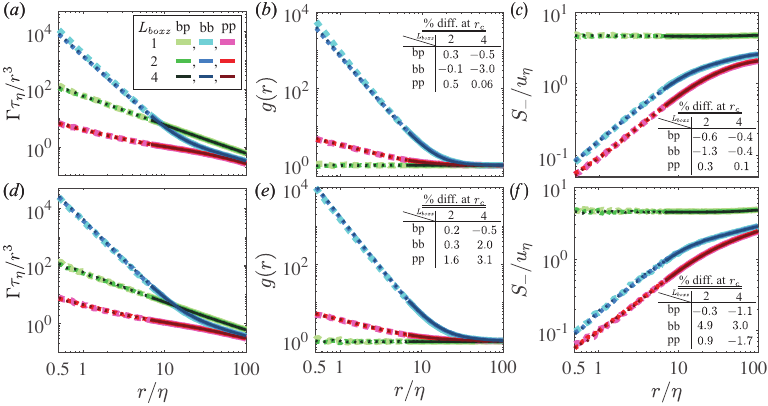}
    \caption{The (\textit{a}) collision kernel, (\textit{b}) RDF and (\textit{c}) effective radial approach velocity with different $L_{boxz}$ at $\Rey_\lambda = 69$ and (\textit{d--f}) $\Rey_\lambda = 167$. The dotted segments indicate $r < r_c$. The tables in (\textit{b--c}) and (\textit{e--f}) show the percentage difference of the values at $r_c$ compared to the $L_{boxz} = 1$ case.}
    \label{fig:periodicityCheck_G}
\end{figure}

The simulations reported are conducted in a cubic domain with $L_{box} = 1$. With gravity, bubble and particle statistics may exhibit periodicity effects if the time taken by the bubbles and particles to travel through the periodic simulation domain is less than the eddy turnover time, i.e. $T_{box,i} < \ell/u'$ \citep{woittiez_combined_2009},  where $\ell$ is the integral length scale. Taking $T_{box,i} \sim L_{boxz}/|\beta_i|\tau_i\mathfrak{g}$, this means
\begin{equation}    \label{eq:Woittiez_maxSt}
    \max(St_i) = Fr\cdot \frac{L_{boxz}u'}{\ell u_\eta|\beta_i|},
\end{equation}
where $L_{boxz}$ is the vertical domain size. Table \ref{tab:maxStInDomain} shows that the maximum $St$ is not exceeded with $L_{boxz} = 1$ even for the largest $St_b$ and $St_p$ simulated except for the $(\Rey_\lambda,1/Fr) = (69,10)$ case. To confirm the collision statistics are not affected even then, we lengthen the simulation domain along the vertical direction following \citet{chouippe_forcing_2015} and rerun the $(St,1/Fr) = (3,10)$ cases. Figure \ref{fig:periodicityCheck_G} shows that $\Gamma(r)$, $g(r)$ and $S_-(r)$ are not sensitive to $L_{boxz}$ for bubble--particle, bubble--bubble and particle--particle collisions.

\begin{table}
  \begin{center}
  \begin{tabular}{lcccccccccccccc}
      {} & \multicolumn{7}{c}{$\Rey_\lambda = 69$}  & \multicolumn{7}{c}{$\Rey_\lambda = 167$}  \\
      \cmidrule(lr){2-8} \cmidrule(lr){9-15}
      $1/Fr$  & 0.01   &   0.1 & 1 & 2 & 3 & 5 & 10 & 0.01   &   0.1 & 1 & 2 & 3 & 5 & 10\\[3pt]
       $\max(St_b)$ & 1947.2 & 194.7 & 19.5 & 9.7 & 6.5 & 3.9 & 1.9 & 3217.7 & 321.8 & 32.2 & 16.1 & 10.7 & 6.4 & 3.2\\
       $\max(St_p)$ & 5338.5 & 533.9 & 53.4 & 26.7 & 17.8 & 10.7 & 5.3 & 8821.7 & 882.2 & 88.2 & 44.1 & 29.4 & 17.6 & 8.8\\
  \end{tabular}
  \caption{The maximum $St$ that satisfies the criterion in (\ref{eq:Woittiez_maxSt}) when $L_{boxz} = 1$.}
  \label{tab:maxStInDomain}
  \end{center}
\end{table}

\section{Code verification}
\begin{figure}
    \centering
    \includegraphics{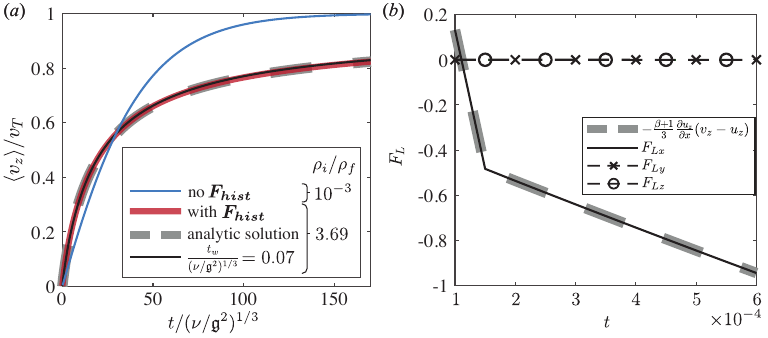}
    \caption{{(\textit{a}) Time series of the mean vertical velocity of 100 bubbles rising in quiescent liquid (blue line) with a Galileo number $Ga_b = \sqrt{\mathfrak{g}(2r_b)^3|\rho_b/\rho_f - 1|}/\nu = 106$, $\rho_b/\rho_f = 1/1000$
    and nonlinear drag $f_b = 1 + 0.169Re_b^{2/3}$, as well as that of a particle settling in quiescent liquid with $Ga_p = 36$, $\rho_p/\rho_f = 3.69$
    and Stokes drag $f_p = 1$.} (\textit{b}) The lift force acting on a bubble with {$(Ga_b,\rho_b/\rho_f) = (106, 1/1000)$} and nonlinear drag $f_b = 1 + 0.169Re_b^{2/3}$ in a simple shear flow.}
    \label{fig:BuoyancyLiftVerification}
\end{figure}

The code used in this study is identical to the already verified code used in \citet{chan_bubbleparticle_2023} apart from the addition of the buoyancy{, history force, and lift terms. To verify the buoyancy term, we first omit the history force and simulate 100 bubbles rising in still fluid. Figure \ref{fig:BuoyancyLiftVerification}(\textit{a}) shows that their terminal rise velocities correspond to the theoretical value ($v_T=0.117$). We next add history force and consider a particle settling in still fluid. Here, we employ Stokes drag ($f_p = 1$) to enable comparison with the analytical solution of the velocity time series by \citet{van_hinsberg_efficient_2011}. As shown in figure \ref{fig:BuoyancyLiftVerification}(\textit{a}), the numerical and analytic solutions agree perfectly when the window length of the history force spans the entire simulation duration, and the agreement is still excellent if a window length of $t_w/(\nu/\mathfrak{g}^2)^{1/3} = 0.07$, which corresponds to 5 time steps, is used. } For the lift term, we checked that the lift force acting on a bubble in a simple shear flow with $\partial u_z/\partial x = 0.64$ is equal to the manually computed value given by the lift force expression {in figure \ref{fig:BuoyancyLiftVerification}(\textit{b})} .

\bibliographystyle{jfm}
\bibliography{BuPaCT_PointParticleSimWithoutGravity}

\end{document}